\documentclass[10pt,twocolumn]{revtex4}
\pdfoutput=1
\usepackage{amsmath}
\usepackage{amssymb}
\usepackage{graphicx}
\renewcommand{\small}[1][]{#1}

\newcommand{\hide}[1]{\relax}
\newcommand{\unit}[1]{\ensuremath{\,\mathrm{#1}}}
\newcommand{\Og}{\ensuremath{\Omega}}
\newcommand{\bD}{\ensuremath{\bar \Delta}}
\newcommand{\meff}{m_\mathrm{eff}}
\newcommand{\xzpf}{\ensuremath{x_{\mathrm{ZPF}}}}
\newcommand{\dwdx}{\ensuremath{g_0}}

\newcommand{\ba}{\ensuremath{\bar a}}

\newcommand{\Om}{\ensuremath{\Omega_\mathrm{m}}}
\newcommand{\Oc}{\ensuremath{\Omega_\mathrm{c}}}
\newcommand{\gzero}{\ensuremath{g_\mathrm{0}}}
\newcommand{\Gm}{\ensuremath{\Gamma_\mathrm{m}}}
\newcommand{\nbarm}{\ensuremath{\bar{n}_\mathrm{m}}}
\newcommand{\nbarmin}{\ensuremath{\bar{n}_\mathrm{min}}}

\newcommand{\SI}{appendix}

\usepackage{array}

\newcommand{\oL}{\ensuremath{\omega_\mathrm{l}}}
\newcommand{\oC}{\ensuremath{\omega_\mathrm{c}}}

\newcommand{\bsin}{\ensuremath{\bar s_\mathrm{in}}}

\newcommand{\mdagger}{\dagger}
\newcommand{\mhat}{}
\newcommand{\mvec}{\vec}%{\mathbf}%{\vec}

\newcommand{\ha}{\ensuremath{\mhat a}}
\newcommand{\hmom}{\ensuremath{\mhat p}}
\newcommand{\hq}{\ensuremath{\mhat q}}
\newcommand{\had}{\ensuremath{\mhat a^\dagger}}

\newcommand{\dha}{\ensuremath{\delta \mhat a}}
\newcommand{\dhad}{\ensuremath{\delta \mhat a^{\mdagger}}}

\newcommand{\dhq}{\ensuremath{\delta \mhat q}}

\newcommand{\dhsin}{\ensuremath{\delta \mhat s_\mathrm{in}}}

\newcommand{\dhsvac}{\ensuremath{\delta \mhat s_\mathrm{cav}}}
\newcommand{\dhsvacd}{\ensuremath{\delta \mhat s_\mathrm{cav}^{\mdagger}}}

\newcommand{\dhFth}{\ensuremath{\delta\! \mhat f_\mathrm{th}}}

\newcommand{\hp}{\ensuremath{\mhat p}}

\newcommand{\vcr}{\ensuremath{g_0}}

\newcommand{\kin}{\ensuremath{\kappa_0}}
\newcommand{\kex}{\ensuremath{\kappa_\mathrm{ex}}}

\begin{document}

\title{Quantum-coherent coupling of a mechanical oscillator\\to an optical cavity mode}
\author{E.\ Verhagen$^{1,\dagger}$, S.\ Del{\'e}glise$^{1,\dagger}$, S.\ Weis$^{1,2,\dagger}$, A. Schliesser$^{1,2,\dagger}$, T.\ J.\ Kippenberg$^{1,2}$}
\email{tobias.kippenberg@epfl.ch}
\affiliation{$^{1}$Ecole Polytechnique F$\acute{e}$d$\acute{e}$rale de Lausanne (EPFL), 1015 Lausanne, Switzerland}
\affiliation{$^{2}$Max Planck Institute of Quantum Optics, 85748 Garching, Germany}
\affiliation{$^{\dagger}$These authors contributed equally to this work.}

\begin{abstract}
Quantum control of engineered mechanical oscillators can be achieved by coupling the oscillator to an auxiliary degree of freedom, provided that the coherent rate of energy exchange exceeds the decoherence rate of each of the two sub-systems. We achieve such quantum-coherent coupling between the mechanical and optical modes of a micro-optomechanical system. Simultaneously, the mechanical oscillator is cooled to an average occupancy of n=1.7$\pm$0.1 motional quanta. Pulsed optical excitation reveals the exchange of energy between the optical light field and the micromechanical oscillator in the time domain at the level of less than one quantum on average. These results provide a route towards the realization of efficient quantum interfaces between mechanical oscillators and optical fields.
\end{abstract}

\maketitle

Mechanical oscillators are at the heart of many precision experiments, such as single spin detection \cite{Rugar2004} or atomic force microscopy and can exhibit exceptionally low dissipation. The possibility to control the quantum states of such engineered micro- or nanomechanical oscillators, similar to the control achieved over the motion of trapped ions \cite{Leibfried2003}, has been a subject of longstanding interest \cite{Braginsky1992, Schwab2005}, with prospects of quantum state transfer \cite{Zhang2003,Tian2010,Akram2010,Stannigel2010}, entanglement of mechanical oscillators \cite{Vitali2007} and testing of quantum theory in macroscopic systems \cite{Marshall2003,Paternostro2007}. However, such experiments require coupling the mechanical oscillator to an auxiliary system\textemdash whose quantum state can be controlled and measured\textemdash with a coherent coupling rate that exceeds the decoherence rate of each of the subsystems. Equivalent control of atoms has been achieved in the context of cavity Quantum Electrodynamics (cQED \cite{Kimble1998}) and has over the past decades been extended to various other systems such as superconducting circuits \cite{Chiorescu2004}, solid state emitters\cite{Hennessy2007} or the light field itself \cite{Deleglise2008}.

Recently, elementary quantum control at the single-phonon level has been demonstrated for the first time, by coupling a piezo-electrical dilatation oscillator to a superconducting qubit \cite{OConnell2010}. An alternative and highly versatile route is to use the radiation-pressure-induced coupling of optical and mechanical degrees of freedom, inherent to optical microresonators \cite{Kippenberg2005}, which can be engineered in numerous forms at the micro- or nanoscale \cite{Kippenberg2008,Clerk2010,Favero2009a}. This coupling can be described by the interaction Hamiltonian $H=\hbar \gzero \hat{a}^\dagger \hat{a} (\hat{b}^\dagger + \hat{b})$, where $\hat{a}$ ($\hat{b}$) is the photon (phonon) annihilation operator, $\hbar$ is the reduced Planck constant and  $\gzero$ is the vacuum optomechanical coupling rate. In the resolved sideband regime (where the mechanical resonance frequency $\Om$ exceeds the cavity energy decay rate $\kappa$), with an intense laser tuned close to the lower optomechanical sideband, one obtains in the rotating wave approximation the effective Hamiltonian
\begin{equation}
	H=\hbar g \left(\hat{a}\hat{b}^\dagger+\hat{a}^\dagger\hat{b}\right)
	\label{effH}
\end{equation}
for the operators $\hat{a}$ and $\hat{b}$ now displaced by their steady state values. We have introduced here the field-enhanced coupling rate \cite{Dobrindt2008,Marquardt2007} $g=\sqrt{\bar{n}_\mathrm{c}}\gzero$, where $\bar{n}_\mathrm{c}$ denotes the average number of photons in the cavity. In the absence of decoherence, the unitary evolution (\ref{effH}) corresponds to swapping of the (displaced) optical and mechanical quantum states with a period of $2 \pi/\Oc$, where $\Oc=2g$ is the coherent energy exchange rate. This state swapping is at the heart of most quantum control protocols \cite{Zhang2003,Tian2010,Akram2010,Stannigel2010}. In practice, however, this unitary evolution is compromised by the coupling of both degrees of freedom to their respective environments. Hence, it is important for $\Oc$ to exceed both the optical decoherence rate  $\kappa$ and the mechanical decoherence rate $\gamma$, defined here as the inverse time needed for a single excitation to be lost into the environment. Reaching this regime of quantum-coherent coupling has proven challenging. While the onset of normal mode splitting \cite{Dobrindt2008, Marquardt2007} (which occurs for $\Oc>\kappa/2$,  when $\kappa$ largely exceeds the mechanical dissipation rate $\Gm$) could be observed in a room temperature experiment \cite{Groblacher2009a}, the mechanical decoherence rate $\gamma=\Gm (\nbarm+1)\approx\Gm \nbarm$ (where $\nbarm=k_\mathrm{B} T / \hbar \Om$) is strongly enhanced by the coupling to the environment at equilibrium temperature $T$. The condition $\Oc>(\kappa,\gamma)$ is analogous to the strong coupling regime encountered in atomic cavity QED \cite{Kimble1998} with the atomic decay rate being replaced by $\gamma$, and the Rabi frequency by the optomechanical coupling rate $\Oc$. Recently, a superconducting microwave cavity electromechanical system has accessed this regime \cite{Teufel2011, Teufel2011b}. If implemented at optical frequencies, the availability of well-developed quantum optical techniques, and the possibility to transport quantum information through room-temperature fiber links could give access to a range of new applications \cite{Zhang2003,Tian2010,Akram2010,Stannigel2010,Groblacher2009a}.

\begin{figure}
\includegraphics[width=1\linewidth]{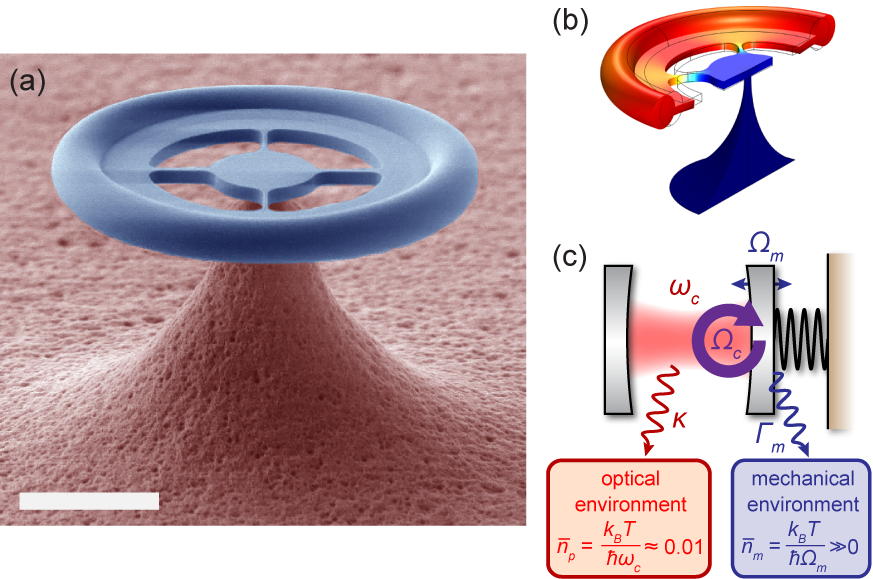}
\caption{
\textbf{Optomechanical resonators.}
(a) Scanning electron micrograph of a spoke-anchored toroidal resonator with 31 $\mu$m diameter used for the optomechanical experiments reported in this work (scale bar 10$\,\mu$m). (b) Simulated displacement (exaggerated for clarity) of the fundamental radial breathing mode of the structure. (c) Equivalent optomechanical Fabry-P{\'e}rot cavity: Quantum-coherent coupling is achieved when the enhanced coupling rate $\Oc$ is comparable to or exceeds the optical and mechanical decoherence rates ($\kappa$, $\Gm \nbarm$). Due to the large asymmetry between mechanical and optical frequencies, the occupations of the two baths are widely different.}
\label{fig1}
\end{figure}

Here we demonstrate for the first time the quantum-coherent coupling of an optical cavity field to a micromechanical oscillator. The experimental setting is a micro-optomechanical system in the form of a spoke-supported toroidal optical microcavity \cite{Anetsberger2008}. Such devices exhibit high quality factor whispering gallery mode resonances (with a typical cavity decay rate $\kappa/2\pi<10\,\mathrm{MHz}$) coupled to mechanical radial breathing modes via radiation pressure. The vacuum optomechanical coupling rate $\gzero=\frac{\omega}{R}x_{\mathrm{ZPM}}$ can be increased by reducing the radius $R$ of the cavity. However, the larger per photon force $\hbar \omega / R$ is then usually partially compensated by the increase in the mechanical resonance frequency $\Om$\textemdash and correspondingly smaller zero point motion $x_\mathrm{ZPM}=\sqrt{\hbar/(2\meff\Om)}$. Moreover, small structures also generally feature larger dissipation through clamping losses. To compensate these opposing effects we use an optimized spoke anchor design (cf. Fig. \ref{fig1} and \SI) that maintains low clamping losses and a moderate mechanical resonance frequency while reducing the dimensions of the structure. Devices fabricated in this manner (with $R=15\,\mu\mathrm{m}$) exhibited coupling rates as high as $\gzero=2\pi \times 3.4\,\mathrm{kHz}$ for a resonance frequency of 78 MHz, as determined independently at room temperature \cite{Gorodetsky2010}.

{\begin{figure}
\begin{center}
\includegraphics[width=1\linewidth]{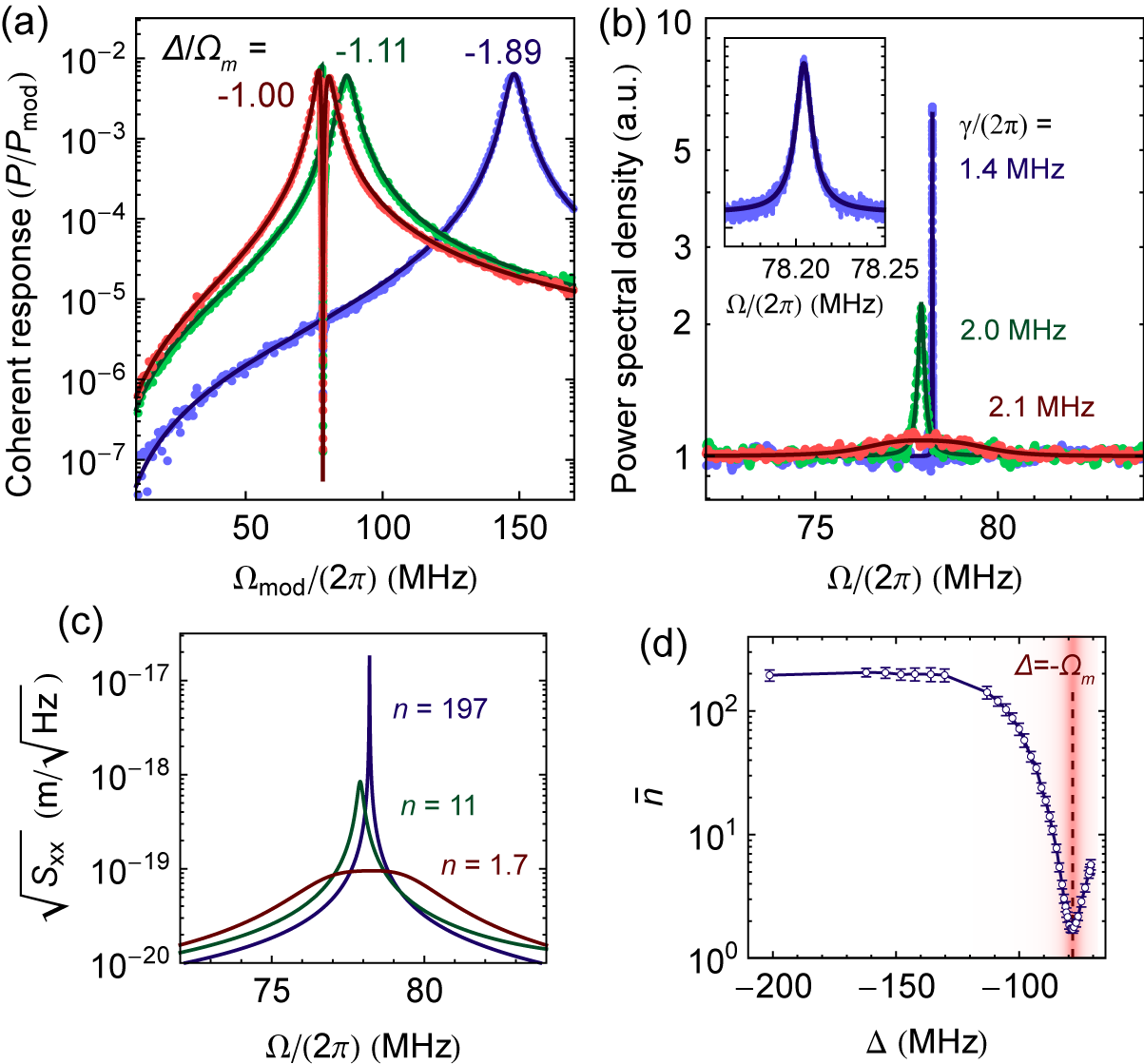}
\end{center}
\caption{
\textbf{Optomechanical interaction in the weak coupling regime ($\Oc\lesssim\kappa/2$).} (a) The coherent response of the system is obtained by sweeping a weak probe beam (created via phase modulation of the coupling laser at frequency $\Omega_\mathrm{mod}$) over the cavity resonance and recording the homodyne signal. The fitted detuning is indicated for each of the traces. The fit provides accurate estimation of the coupling rate $\Oc$ via the Optomechanically Induced Transparency window (see \SI$\,$  for details). (b) The measured Brownian noise spectrum in the absence of a coherent probe for each of the detunings in (a). The spectra are corrected for the detector response and a small contribution of Guided Acoustic Wave Brillouin scattering in the fibers (see \SI). The inset shows a close-up of the spectrum obtained for $\Delta/\Om=-1.89$. The indicated decoherence rates $\gamma$ are deduced from the amplitude of each noise spectrum. (c) Inferred mechanical displacement spectra calculated using the extracted parameters. (d) Retrieved occupancy as a function of detuning.}%
\label{fig2}
\end{figure}}

{\begin{figure*}
\begin{center}
\includegraphics[width=0.9\linewidth]{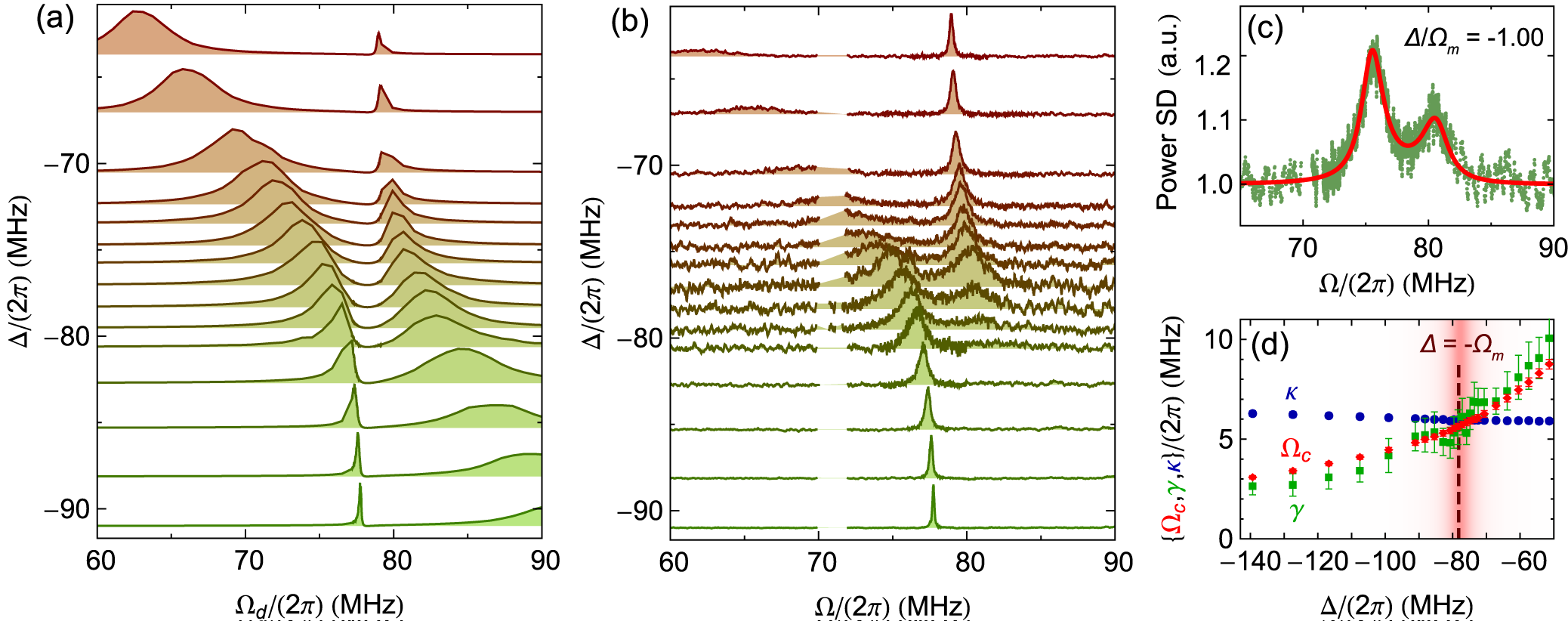}
\end{center}
\caption{
\textbf{Quantum-coherent coupling.} The coherent optical response (a) and the incoherent mechanical noise spectrum (b) for various detunings of the coupling laser (with constant power P = 1.4 mW). All curves are normalized, and vertically displaced by the detuning. Evident in both panels is the avoided crossing which originates from optomechanical normal mode splitting. A second much smaller mechanical mode at 71 MHz is omitted. (c) Homodyne noise spectrum obtained for $\Delta=-\Om$. The red line is a fit of the model. Only the decoherence rate (determining the amplitude) is fitted, while the shape is fixed through the parameters determined from the coherent response. (d) Comparison of the mechanical (green) and optical (blue) decoherence rates with the coherent coupling rate (red) as a function of detuning. The increase of $\gamma$ close to resonance reflects heating of the cavity due to the larger amount of absorbed light. On the lower mechanical sideband, where the interaction is resonant, the decoherence rates are comparable to the coupling rate, achieving quantum-coherent coupling.}
\label{setup}%
\label{fig3}
\end{figure*}}

\begin{figure}[t]
\begin{center}
{\includegraphics[width=0.95\linewidth]{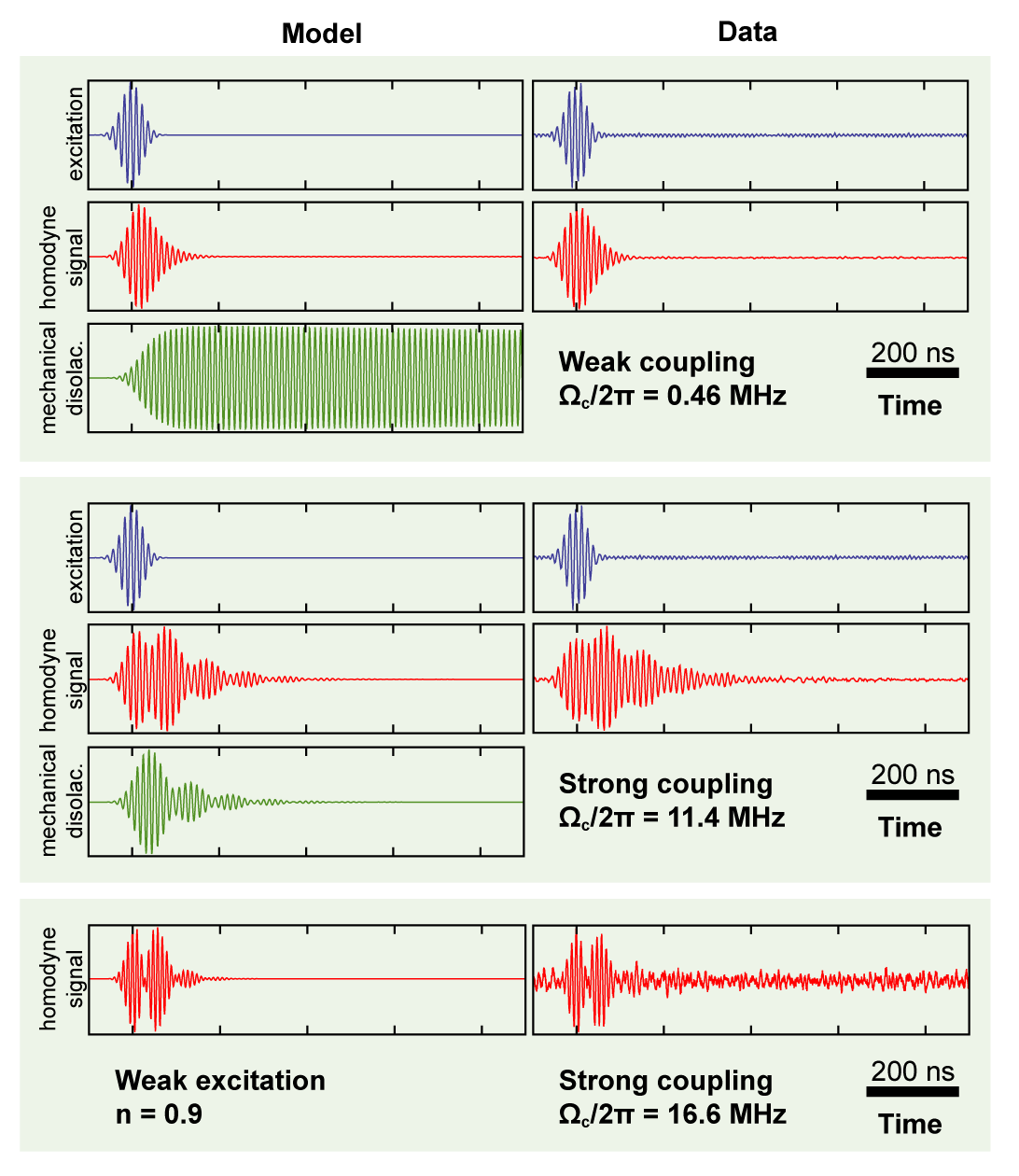}  }
\end{center}
\caption{
\textbf{Coherent exchange between the optical field and the micromechanical oscillator} probed in the time domain as measured (``Data'') and calculated numerically (``Model''). A modulation pulse (blue traces) applied to the phase modulator creates a light pulse probing the dynamics of the optomechanical system. The response of the system is encoded into the optical pulse at the output of the coupling fiber as recorded by the homodyne receiver (red traces, 250,000 averages). Using the full model of the system, the mechanical displacement can be simulated in addition (green traces). In the regime of weak coupling (top panel), the optical output pulse exhibits only a weak signature of the nearly unperturbed mechanical ringdown excited by the short burst of radiation pressure. In the case of strong coupling (middle and bottom panel) the modulated envelopes of the time-domain response indicate several cycles of oscillation between (very-low energy $\bar{n}=0.9$, bottom) coherent optical and mechanical excitations of the system.}
\label{fig4}
\end{figure}

To reduce the mechanical decoherence rate $\gamma=\Gm\nbarm$, the microcavity, coupled to a tapered optical fiber, is embedded in a helium-3 cryostat \cite{Riviere2011}. For low optical power, the $^\mathrm{3}\mathrm{He}$ buffer gas enables thermalization of the resonator over the entire cryogenic temperature range ($T_\mathrm{min}=650\,\mathrm{mK}$) in spite of its weak thermal anchoring to the substrate. A quantum-limited continuous-wave Ti:Sapphire laser provides both the coupling field and homodyne local oscillator used to measure the weak phase fluctuations imprinted on the field emerging from the cryostat by mechanical displacement fluctuations. While the coherent coupling rate $\Omega_c$  can be determined unambiguously by probing the coherent response of the system \cite{Weis2010}, the mechanical decoherence rate is affected in a non-trivial way by  the light-absorption-dependent sample temperature and the mechanical mode's coupling to its environment \cite{Riviere2011}, which is dominated by two-level fluctuators at cryogenic temperatures \cite{Arcizet2009a}. In order to systematically assess the decoherence rate, the coupling laser's frequency $\omega_\mathrm{l}=\omega_\mathrm{c}+\Delta$ is swept in the vicinity of the lower mechanical sideband. This allows to bring the  displaced cavity mode $\hat{a}$  (of frequency $\left|\Delta\right|$) in and out of resonance with the mechanical mode $\hat{b}$ (of frequency $\Omega_m$). For each detuning point, we acquire the coherent response of the system to an optical excitation in a first step using a balanced homodyne detection scheme. These spectra (Fig \ref{fig2}a) allow to determine all parameters of the model characterizing the optomechanical interaction (\SI). For large detunings $\left|\Delta\right|>\Om$, they essentially feature a Lorentzian response of width $\kappa$ and center frequency $\left|\Delta\right|$. The sharp dip at $\Omega\approx\Om$ originates from Optomechanically Induced Transparency \cite{Weis2010}, and for $\Om=-\Delta$, its width is approximately $\Oc^2/\kappa$. The coupling rate, as derived from a fit of the coherent response for a laser power of 0.56 mW, is $\Oc=2\sqrt{\bar{n}_\mathrm{c}}\gzero=2\pi \times(3.7\pm0.05)\,\mathrm{MHz}$  (corresponding to an intracavity photon number of $\bar{n}_\mathrm{c}=3\cdot10^5$).

Additionally, for each value of the detuning the noise spectrum of the homodyne signal is recorded in the absence of any external excitation (Fig. \ref{fig2}b). The observed peak represents the phase fluctuations imprinted on the transmitted light by the mechanical mode's thermal motion. The constant noise background on these spectra is the shot-noise level for the (constant) laser power used throughout the laser sweep (see \SI $\;$ for details). The amplitude of the peak is determined by the coupling to and the temperature of the environment, and therefore allows to extract the mechanical decoherence rate. All parameters now being fixed, it is moreover possible to retrieve the mechanical displacement spectrum (Fig. \ref{fig2}c). As can be seen, for detunings close to the sideband, when the (displaced) optical and mechanical modes are degenerate, the fluctuations are strongly reduced. This effect of optomechanical cooling can be understood in a simple picture: In the regime $\Oc\ll\kappa$, the optical decay is faster than the swapping between the vacuum in the displaced optical field and the thermal state in the mechanical oscillator. In this case, the mechanical oscillator is coupled to an effective optical bath at near-zero thermal occupancy  $\nbarmin$ with the rate $\Gamma_\mathrm{cool}=\Oc^2/\kappa$. Ideally, $\nbarmin=\kappa^2/16\Om^2\ll1$ is governed by non-resonant Stokes terms $\hat{a}^\dagger\hat{b}^\dagger+\hat{a}\hat{b}$ neglected in the Hamiltonian (\ref{effH}). Any excess noise in the coupling beam will, however, cause an effective increase of $\nbarmin$ and precludes any quantum state manipulation due to the impurity of the field's quantum state. In practice it has proven crucial to eliminate phase noise originating from Guided Acoustic Wave Brillouin Scattering \cite{Shelby1985a} in the optical fibers by engineering their acoustic modes using HF etching (\SI).

Evaluating the mechanical decoherence rate for $\Delta=-\Om$ at a cryostat setpoint of 0.65~K, we find $\gamma=2 \pi \times (2.2\pm0.2)\,\mathrm{MHz}$\textemdash significantly smaller than $\Oc$. Simultaneously, the average occupancy of the mechanical mode is reduced to $\bar{n}=1.7\pm0.1$, which is limited by the onset of normal mode splitting. Indeed, as $\Oc$ approaches $\kappa$, the thermal fluctuations are only partially dissipated into the optical bath, and partially written back onto the mechanics after one Rabi cycle. Note that this occupancy, corresponding to 37\% ground state occupation, is associated with a sideband asymmetry of $1-\bar{n}/(\bar{n}+1)\approx 40\%$. A measurement of this asymmetry, similar to those performed on trapped ions \cite{Leibfried2003}, would yield direct evidence of the quantum nature of macroscopic mechanical oscillators.

We subsequently increase the strength of the coupling field to reach $\Oc\approx\kappa$. The signature of normal mode splitting \cite{Dobrindt2008,Marquardt2007,Groblacher2009a,Teufel2011}  can be seen from both the coherent response and the fluctuation spectra in Fig.~\ref{fig3}. Both detuning series exhibit a clear anti-crossing, the splitting frequency being $5.7\,\mathrm{MHz}$. The decoherence rate (see Fig.~\ref{fig3}d) is slightly raised compared to Fig.~\ref{fig2} due to laser heating and a higher buffer gas temperature of 0.8K, amounting to $\gamma=2\pi\times (5.6\pm0.9)\,\mathrm{MHz}$ at the lower mechanical sideband. We hence demonstrate $\Oc/\gamma=1.0$, which constitutes a four-orders of magnitude improvement over previous work in the optical domain \cite{Groblacher2009a}, and brings the system into the regime of quantum-coherent coupling.

As a proof of principle, we finally demonstrate the dynamical exchange of weak coherent excitations between the optical and mechanical degrees of freedom in the time domain (Fig.~\ref{fig4}). To this end, we modulate the coupling laser's phase at the mechanical resonance frequency, with a Gaussian envelope of 54 nano-second duration. This modulation creates a pair of sidebands\textemdash one of which is resonant with the optical cavity\textemdash that contain on average ten quanta per pulse (cf. \SI). By measuring the homodyne signal it is possible to directly observe the coherent exchange of energy.

In the regime of weak coupling (Fig.~\ref{fig4}a) the optical pulse excites the mechanical mode to a finite oscillation amplitude, which decays slowly with the nearly unperturbed mechanical dissipation rate. This in turn only imprints a weak signature on the homodyne signal. Increasing the laser power, we reach $\Oc / 2 \pi = 11.4 \mathrm{MHz}>\kappa / 2 \pi=7.1\mathrm{MHz}$, and the envelope of the homodyne signal\textemdash corresponding to the amplitude of the measured sideband field\textemdash undergoes several cycles of energy exchange with the mechanical oscillator, before it decays with a modified rate of $\left(\kappa+\Gm\right)/2\approx\kappa/2$, corresponding to the decay rate of the optomechanical polariton excited. Using our model matched to previously taken coherent response measurements, we can derive not only the homodyne signature\textemdash which reproduces our data very well\textemdash but also the expected mechanical oscillations resulting from this pulsed excitation (see Fig.~\ref{fig4}). These reveal very clearly how the excitation cycles continuously between the optical and mechanical modes. Although the employed detection is far from optimized for time-domain experiments, we have achieved a signal-to-noise ratio of 40 by averaging 250,000 traces within a total acquisition time of approximately two minutes. Reducing the excitation further so that one pulse contains less than one photon on average, a clear signature of several swapping cycles can still be observed (Fig.~\ref{fig4}c). Replacing the weak coherent state by a single photon \cite{Akram2010}, we expect Rabi-like oscillation of the Fock state from optics to mechanics and back. A repeated quadrature measurement yielding a bimodal distribution would be an unambiguous signature of the quantum nature of the state after the full swap \cite{Lvovsky2001}.

Obtaining quantum-coherent coupling $\Oc\gtrsim(\gamma,\kappa)$ has several interesting consequences. First, it allows the mapping of in principle any quantum state of the optical field onto the mechanical mode via the use of a time-dependent coupling field. As a simple example, initialization of the mode in the ground state can be efficiently achieved in this regime using a  \mbox{$\pi$-pulse} which swaps the thermal state of the oscillator and the vacuum in the optical field \cite{Jacobs2010}. Note that the manipulation of large quantum states becomes increasingly challenging since the lifetime of the number state $|n\rangle$ scales with $1/n$. In this context, it will be beneficial to further reduce the decoherence rate by limiting spurious laser heating and employing materials with low intrinsic loss, as well as to increase the optomechanical coupling rate by further miniaturization. The regime of quantum-coherent coupling demonstrated here has been proposed as a general quantum link between electromagnetic fields of vastly different frequencies, e.g., different wavelengths in the optical spectrum or microwave and optical photons \cite{Tian2010,Regal2011}. In that respect, the efficient coupling of the demonstrated system to a low-loss single mode optical fiber is beneficial. Moreover, quantum-coherent coupling enables the use of the mechanical oscillator as a transducer to link otherwise incompatible elements in hybrid quantum systems, such as solid-state spin, charge, or superconducting qubits and propagating optical fields \cite{Stannigel2010}. In this context, we note that the mechanical modes of silica microresonators have already been shown to be intrinsically coupled to effective two-level systems in the form of structural defect states in the glass \cite{Arcizet2009a}.

In conclusion, the reported experiments\textemdash demonstrating quantum coherent coupling between a micromechanical oscillator and an optical mode\textemdash represent a first step into the experimental investigation and quantum optical control of the most tangible harmonic oscillator: a mechanical vibration.

{\small $\hphantom{xxx}$\newline\noindent\textbf{Acknowledgements}
 This work was supported by the ERC Starting Grant SiMP, The Defense Advanced Research Agency (DARPA), The NCCR of Quantum Engineering and the SNF. E.V. acknowledges support from a Rubicon Grant by NWO, cofinanced by a Marie Curie Cofund Action. S.D. is funded by a Marie Curie Individual Fellowship.\newline}

{
\newcommand{\nocontentsline}[3]{}
\renewcommand{\addcontentsline}[2][]{\nocontentsline#1{#2}}

}

\clearpage

%%%%%%%%%%%%%%%%%%%%%%%%%%%%%%%%%%%%%%%%%%%%%%%%%%%%%%%%%%%%%%%%%%%%%%%%%%%%%%%%%%%%%

\onecolumngrid

\setcounter{figure}{0}%
\setcounter{equation}{0}%
\setcounter{section}{0}
\renewcommand \theequation {A\arabic{equation}}%
\renewcommand \thefigure {A\arabic{figure}}
\renewcommand \thesection {A\arabic{section}}

\begin{center}
%\vspace{0.5in}
\Large{
\textbf{Appendix}}
\end{center}
\vspace{-0.2in}

\setcounter{tocdepth}{3}
\tableofcontents

\section{Experimental details}

\subsection{Experimental setup}
\label{ss:setup}

\begin{figure}
	\centering
		\includegraphics[width=0.65\linewidth]{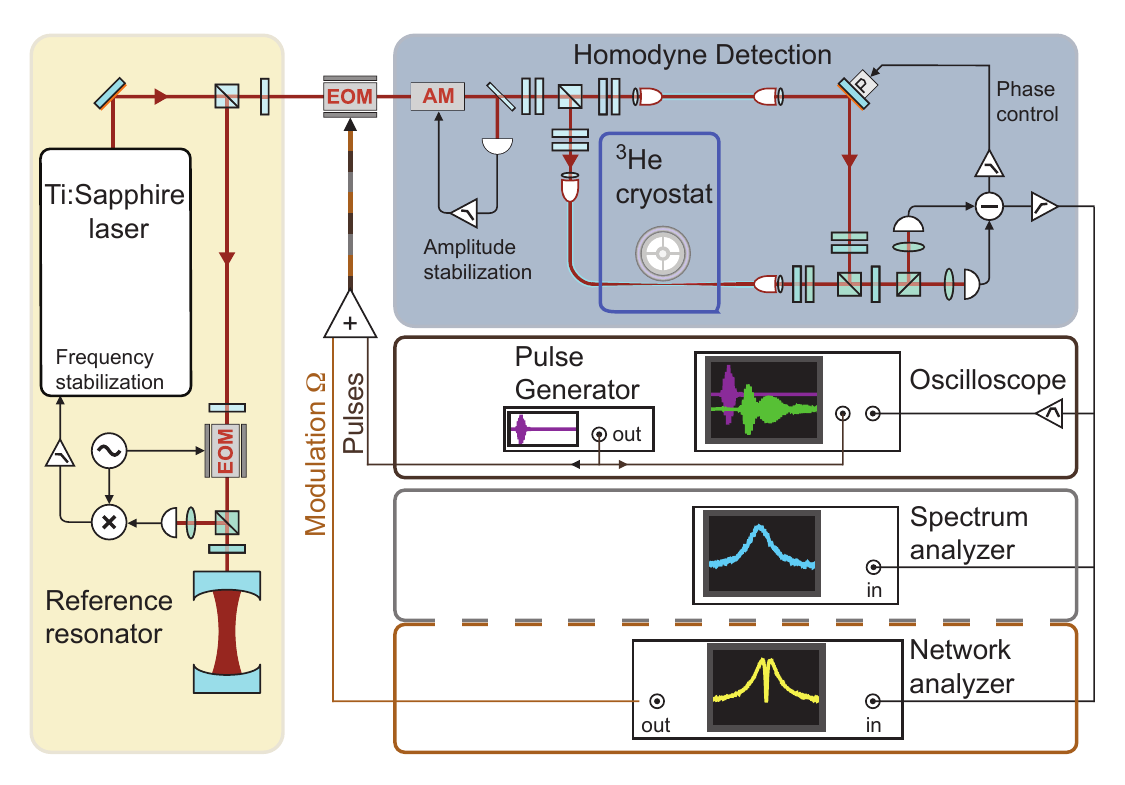}
	\caption{\textbf{Setup.} See text for details.}
	\label{fig:Setup}
\end{figure}

Figure \ref{fig:Setup} shows a schematic of the employed experimental setup. At the heart of the optical setup is a Ti:Sapphire laser (Sirah Matisse TX) operating at a wavelength around $780\,\mathrm{nm}$. The laser exhibits quantum limited amplitude and phase noise at Fourier frequencies relevant for this experiment. During the experiments the laser is locked to an external reference cavity such that drifts of the laser detuning $\Delta$ can be neglected during the acquisition time.

The sample itself resides in a helium-3 exchange gas cryostat (Oxford Instruments Heliox TL) that is used for cryogenic pre-cooling of the mechanical mode to low temperatures. Since the toroids are situated directly above the surface of the liquified helium-3, the achievable temperature is directly linked to the vapor pressure curve of helium-3. As the toroidal microstructures are thermally very well isolated from the substrate, one relies on cooling via the helium-3 exchange gas. As a consequence, cryostat temperature setpoints of at least 650 mK (corresponding to pressures larger than $\approx0.15\,\mathrm{mbar}$) are favorable. Coupling of light into the toroid is achieved via a tapered optical fiber that is approached using piezo positioners, which are compatible with low temperature operation (Attocube GmbH). The fiber ends are guided through and out of the cryostat and constitute one arm of the interferometer that is part of the balanced homodyne detection scheme. The length of one of the ca. 8 m long arms is servo-locked using a movable mirror to cancel the DC-component of the interferometer's signal. This setting allows a shot-noise limited read-out of the phase noise imprinted onto the transmitted laser field. The laser passes an electro-optical modulator that allows to create sidebands around the laser frequency. Last, the laser power is stabilized actively in absolute terms at the input of the experiment to ensure operation at a constant light intensity.

The three colored building blocks highlighted in Figure \ref{fig:Setup} depict the three different measurements that are routinely performed one after the other.

\begin{itemize}
	\item \emph{Coherent response.}	A network analyzer sweeps the upper modulation sideband over the optical resonance and demodulates the corresponding (coherent) signal (cf. section \ref{CoherentDynamics}).\\
	\item \emph{Noise spectrum.} Connecting only an electronic spectrum analyzer gives access to the incoherent noise spectrum (cf. section \ref{NoiseCovariances}).\\
	\item \emph{Time domain response.} Sending a pulsed stimulus from an arbitrary waveform generator to the EOM which modulates the coupling laser gives access to the dynamic time domain response of the optomechanical system (cf. section \ref{timeDomain}).\\
\end{itemize}

\subsection{Influence of guided acoustic wave Brillouin scatting \label{GAWBS}}
A crucial prerequisite for optomechanical measurements in the quantum regime is the use of a quantum limited laser source. From the point of view of quantum manipulations, added noise in the coupling beam corresponds to an improper state preparation, the optical beam being in a statistical mixture of pure quantum states. In the weak coupling limit $\Omega_c \ll \kappa$, where the optical field acts as an effective bath, these extra fluctuations correspond to an increased temperature of the bath and prevents cooling close to the quantum ground state \cite{Schliesser2008B, Rabl2009B, Diosi2008B}. In addition, classical laser noise driving the optomechanical system can lead to ambiguous signatures such as squashing in the noise spectra, as reported previously \cite{Rocheleau2010B}. We have verified in our previous work \cite{Riviere2011B}, that the employed laser source is quantum limited. However, as is well known from fiber-based quantum optics experiments \cite{Shelby1985aB}, optical fibers can give rise to classical phase noise, in the form of guided acoustic wave Brillouin scattering (GAWBS). This process involves thermally driven radial mechanical modes of the fiber, that also modulate the optical path length.

\begin{figure}[b]
	\centering
		\includegraphics[width=0.7\textwidth]{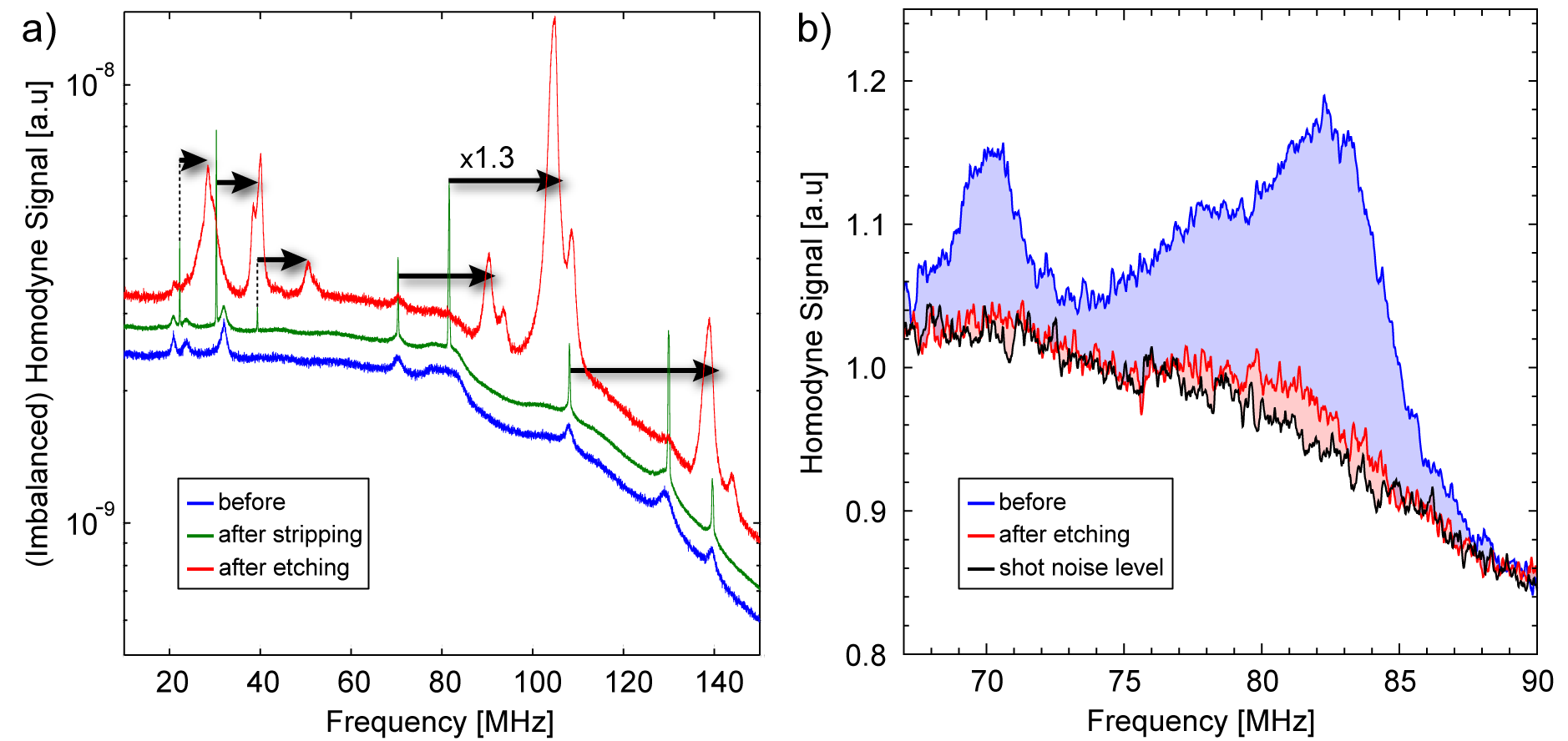}
	\caption{\textbf{Engineering of the fiber GAWBS noise spectrum.} The figure shows a broadband background spectrum of the imbalanced homodyne signal where the GAWBS modes are visible. Blue, green and red trace are taken with unmodified, partly stripped buffer and (almost entirely) etched fiber respectively. As expected for guided dilatational acoustic waves of the optical fiber the frequencies are increased by a factor of about 1.3 for a thinned fiber of around $95\,\mu\mathrm{m}$ diameter (as compared to $125\,\mu\mathrm{m}$ before). Doublets in the red trace are due to slightly different final etching radii (difference is about $3\,\mu\mathrm{m}$) of the different fibers in our setup (i.e., local oscillator fiber and the signal fiber). The difference in relative heights of the peaks is attributed to varying readout conditions, and as such only the peak's frequencies are of interest. The inset shows a zoom of the background for the final setup (i.e., etching reduced diameter fibers, balanced homodyne arm lengths) and illustrates the achieved improvements, i.e., the reduced contribution of GAWBS to the background at the mechanical resonance frequency.}
	\label{fig:GAWBS}
\end{figure}

To investigate the presence of GAWBS we have recorded the noise spectrum from the homodyne detector when the fiber is retracted away from the cavity in an imbalanced Mach-Zehnder interferometer. Several classical peaks are observed on top of the shot-noise background (not corrected for the detector response)(Fig. \ref{fig:GAWBS}).
The width (i.e. damping) of the noise peaks was observed to narrow dramatically when the buffer was partly stripped off the fiber, clearly demonstrating the mechanical nature of the peaks.
One of the peaks coincides with the mechanical resonance frequency of $78\,\mathrm{MHz}$. However, the frequency of the dilatational fiber modes is proportional to the inverse fiber radius and can therefore be shifted by etching the fiber cladding in an HF solution. Immersing the fibers (without removing the acrylate buffer, which is permeable to HF) in a $40\,\%$ HF solution for 50 minutes reduced the cladding diameter from $125\,\mu\mathrm{m}$ to $95\,\mu\mathrm{m}$. This increased the GAWBS mode frequencies of all fibers in the setup by $\approx30\,\%$, shifting them away from the mechanical resonance frequency of the toroid (Fig. \ref{fig:GAWBS}a).

The data shown in Fig. \ref{fig:GAWBS}b have been taken under the same conditions as the lowest occupation run shown in Fig. 2 of the main manuscript. The residual noise at $78\,\mathrm{MHz}$ (due to small portions of fibers that have not been etched) is approximately a factor seven smaller than the initial peaks, corresponding to a noise level of approximately $2\,\%$ of the shot-noise.
This noise is generated along the fibers, both before and after the cavity. Special care was taken to minimize the length of unetched fiber before the cavity. The fact that an influence of the cavity detuning and coupling parameters on the transduction of these classical noise peaks into a measured signal is not discernible indicates that indeed most remaining noise originates from fiber after the cavity. Under this assumption, the independent noise of the GAWBS can be subtracted from the signal in order to estimate the decoherence rate and occupation. Figure \ref{fig:GAWBS}c in the main text shows that the shape of the spectra, as predicted from independently measured parameters, is in excellent agreement with the data after subtraction, in which no signs of squashing are observed.
Nonetheless, we have performed an additional analysis for the lowest-occupancy data under the assumption that half of the noise is generated before the cavity, which leads to deviations of the decoherence rate and occupation of $7\,\%$ and $5\,\%$, respectively. This upper bound of the influence of GAWBS, corresponding to an uncertainty of 0.08 phonons, is included in the quoted errors (cf. section \ref{ErrorAnalysis}).

\subsection{Time-domain response \label{timeDomain}}
In order to probe the coherent dynamics of the optomechanical system in the time domain, the strong pump beam is tuned to the red sideband, and an RF pulse, resonant with the mechanical oscillator, is sent to the EOM. The upper modulation sideband excites the strongly coupled system. The subsequent evolution of the transmitted signal is recorded using the homodyne detector and an oscilloscope.

%\begin{itemize}
An arbitrary signal generator (Agilent 33250A) is used to generate the RF pulses. The time dependent voltage $U(t)$ is a sine wave modulated by a Gaussian envelope:
\begin{align}
 U(t)&=E(t) \sin(\Omega_\mathrm{mod} t+\phi_\mathrm{0})\\
E(t)&=U_0 e^{-\left(\frac{t-t_0}{\tau}\right)^2}
\end{align}
with a carrier frequency $\Omega_\mathrm{mod} = 2 \pi \times 77\,\mathrm{MHz}$ and an envelope duration $\tau = 32\,\mathrm{ns}$ ($\mathrm{FWHM} = 54\,\mathrm{ns}$).
A digital oscilloscope, synchronously triggered with the signal generator is used to record and average the homodyne response. The very small signal originating from the balanced detectors is amplified and filtered, around a frequency of $75\,\mathrm{MHz}$, with a bandwidth of $100\,\mathrm{MHz}$.
For low excitation amplitude, averaging is necessary to extract the coherent response out of the incoherent thermal and quantum noises from the optomechanical system.

The modulation depth $\beta(t)$ corresponding to the instantaneous value of the slowly varying envelope is given by:
\begin{align}
 \beta(t)=\pi \frac{E(t)}{V_\pi},
\end{align}
where $V_\pi = 154$ V is the voltage corresponding to  a $\pi$ phase shift of the beam in the EOM (NewFocus 4002). For a weak modulation depth ($\beta \ll 1$), a fraction $\left(\beta/2\right)^2$ of the optical carrier power $P_\mathrm{c}$ is scattered into each of the two first modulation sidebands.
The total optical power in the upper sideband can hence be simply approximated by
\begin{align}
 P(t)\approx P_c \left(\pi \frac{E(t)}{2 V_\pi}\right)^2
\end{align}
The total energy in the pulse can then be obtained by integrating the instantaneous power over the duration of the pulse. The average number of photons in one pulse is hence given by:
\begin{align}
n \approx \frac{1}{\hbar \omega}\int_{-\infty}^{+\infty}{P_c \left(\frac{\pi E(t)}{2 V_\pi}\right)^2 \mathrm{d}t} =  \frac{ \pi^{5/2} }{4 \sqrt{2} } \frac{\tau  P_c}{\hbar \omega} \left(\frac{U_0}{V_\pi}\right)^2
\end{align}

\section{Optimized spoke-anchored toroidal resonator}

\subsection{Sample design \label{SampleDesign}}
\begin{figure}[!b]
	\centering
		\includegraphics[width=\textwidth]{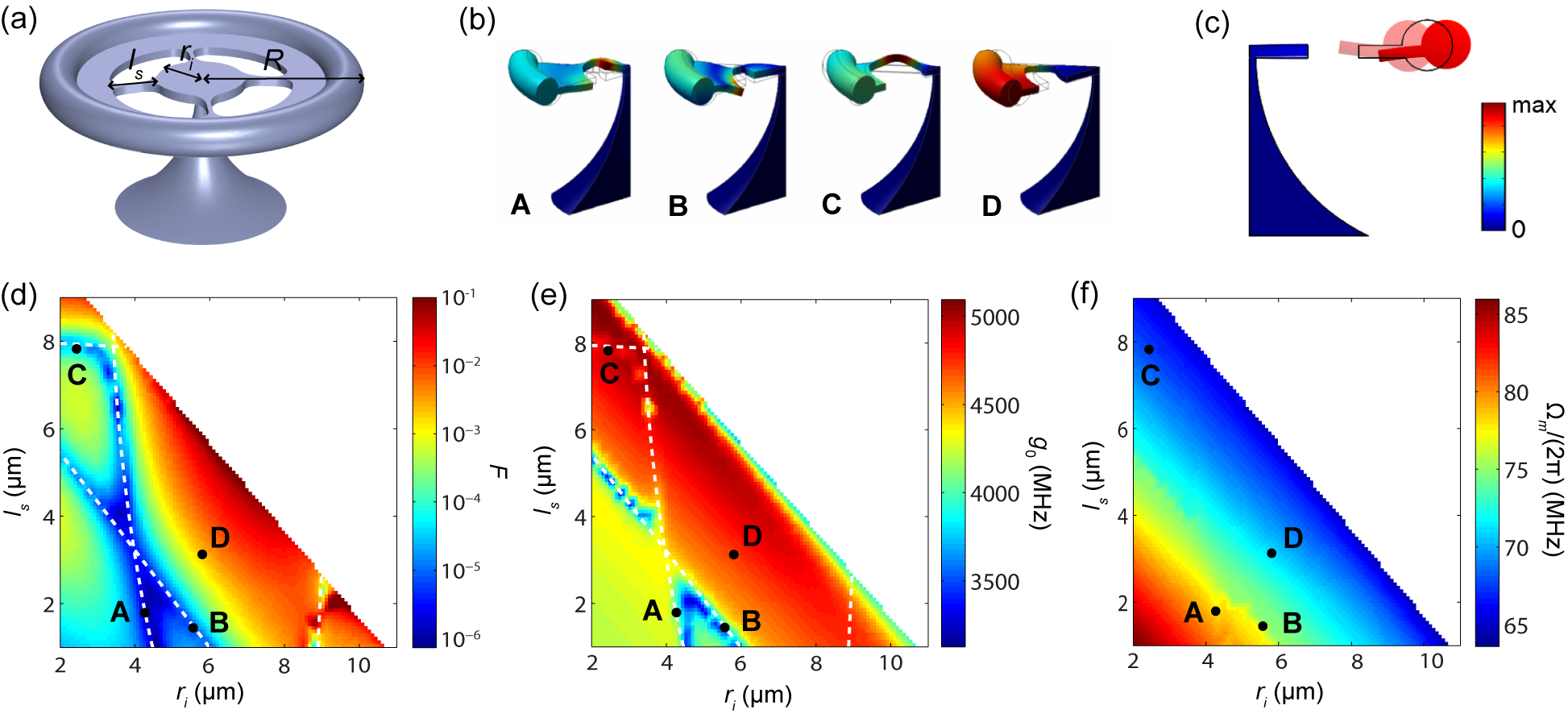}
	\caption{\textbf{Sample optimization by finite element modeling.} See text for details.}
	\label{fig:SampleDesign}
\end{figure}

The optomechanical microresonators investigated in this work are specially designed toroidal microcavities, optimized to achieve large optomechanical coupling rates and small dissipation. Toroidal silica whispering gallery mode microresonators exhibit mechanical modes coupled to the optical modes through radiation pressure \cite{Kippenberg2005B}. Of particular interest is the lowest order radial breathing mode (RBM), whose motion maximally modulates the optical cavity length. In the context of quantum-coherent coupling, it is important to simultaneously achieve large values of $\Omega_\mathrm{c}/\gamma$ and $\Omega_\mathrm{c}/\kappa$, where $\Omega_\mathrm{c}=2 g_\mathrm{0} \left|\bar{a}\right|$. The vacuum optomechanical coupling rate $g_0$ is given by $\frac{\omega}{R}\sqrt{\hbar/\left(2m_\mathrm{eff}\Omega_\mathrm{m}\right)}$, where $R$ is the toroid radius, $m_\mathrm{eff}$ is the effective mass, and $\omega$ and $\Omega_\mathrm{m}$ are the optical and mechanical resonance frequencies, respectively. For a given incident power, optical frequency, and environment temperature, and assuming the resolved sideband regime and the coupling laser being tuned to the lower mechanical sideband, one obtains $\Omega_\mathrm{c}/\gamma \propto \left( R \,\Gamma_\mathrm{m} \sqrt{\Omega_\mathrm{m} m_\mathrm{eff} / \kappa}\right)^{-1}$ and $\Omega_\mathrm{c}/\kappa \propto \left( R \,\Omega_\mathrm{m}^{3/2} \sqrt{m_\mathrm{eff} \kappa}\right)^{-1}$. It is therefore obviously beneficial to reduce the sample dimensions to decrease $R$ and $m_\mathrm{eff}$. However, such miniaturization is generally accompanied by an increase of the mechanical frequency $\Omega_\mathrm{m}$, as well as an increase of $\Gamma_\mathrm{m}$ due to enhanced clamping losses. In microtoroids, both of these adverse effects can be countered in a design in which the toroid is suspended by spokes from the central pillar, as shown in Fig. \ref{fig:SampleDesign}a.

The introduction of spokes serves three purposes. First, they isolate the mechanical motion of the toroidal RBM from the pillar support, strongly reducing clamping losses \cite{Anetsberger2008B}. Second, they reduce the mechanical mode volume and thereby the effective mass. Third, the effective spring constant is reduced, which lowers the mechanical resonance frequency. In practice, one needs to carefully consider the precise dimensions and positioning of the spokes, as these strongly affect both clamping losses $\Gamma_\mathrm{clamp}$ and $g_0$. Figure \ref{fig:SampleDesign}b shows the displacement profile of the RBM of a spoke-supported toroid of radius $R=15$ $\mu$m for various combinations of spoke length and position, as simulated with a finite element method. 
The SiO$_2$ thickness is 1 $\mu$m, the minor toroid radius is 2 $\mu$m, the spoke width is 500 nm, the pillar diameter is 1 $\mu$m, and the toroid is vertically offset from the middle SiO$_2$ disk by 400 nm. Since we are interested in the RBM only, it suffices to simulate 1/8 portion of the microresonator while assuming symmetric boundary conditions on both of the two `cut' planes. As can be seen from these examples, the mechanical mode profiles can change drastically depending on the spoke dimensions. Of the examples in Fig. \ref{fig:SampleDesign}b, only `D' depicts a mode that is purely localized to the outer toroid, with purely radial displacement, as illustrated in the cross-section in Fig. \ref{fig:SampleDesign}c.

The origin of this wildly varying nature of the RBM is revealed in Fig. \ref{fig:SampleDesign}d, where the radius $r_\mathrm{i}$ of the inner disk (defining the spoke placement) and the spoke length $l_\mathrm{s}$ are varied systematically. The colorscale depicts the parameter $F$, defined as
\begin{equation}
F = \frac{2\pi E_\mathrm{mech}}{c \rho \Omega_\mathrm{m}^2 \int_{A_\mathrm{p}} \left| \Delta z \left( \mathbf{x} \right) \right|^2 dA}.
\end{equation}
Here, $E_\mathrm{mech}$ is the total mechanical energy in the mode, $c$ the speed of sound in silica, $\rho$ the density of silica, and $\Delta z\left(\mathbf{x}\right)$ the out-of-plane displacement amplitude, with the integration extending over the area $A_\mathrm{p}$ of the interface between the pillar and the silica disk. $F$ is proportional to the expected value of $\Gamma_\mathrm{clamp}^{-1}$, when the clamping area $A_\mathrm{p}$ is considered as a membrane radiating energy with a power $P=c \rho \Omega_\mathrm{m}^2 \int_{A_\mathrm{p}} \left| \Delta z \left( \mathbf{x} \right) \right|^2 dA$ \cite{Anetsberger2008B}. A previous study has experimentally found a correspondence of $F\approx\left(3\Gamma_\mathrm{clamp}/\left(2\pi\right)\right)^{-1}$ for larger toroids. As can be seen from the figure, the expected clamping losses vary strongly with spoke dimensions, ranging from $10^1$ to $10^6$ Hz. Most notably, several lines can be identified in this parameter space where clamping losses are large (indicated by the dashed lines). For parameter combinations along each of these lines, the RBM frequency approaches that of another mechanical mode of the structure. As a result, the two modes exhibit an anticrossing, with the hybridized modes showing a character of both uncoupled modes. This is the case for examples `A', `B', and `C' in Fig. \ref{fig:SampleDesign}b, which show the RBM hybridized with a flexural mode of the inner SiO$_2$ disk, the outermost SiO$_2$ membrane, and the spoke itself, respectively. In the vicinity of these anticrossings, the vertical displacement at the pillar, and as such the radiation into the substrate $F^{-1}$, are strongly enhanced. To achieve a design that exhibits small clamping losses, it is therefore crucial to avoid these parameter regions, as is the case for mode `D' in Fig. \ref{fig:SampleDesign}b.

The aforementioned anticrossings affect the coupling rate $g_0$ as well, albeit to a lesser degree. Fig. \ref{fig:SampleDesign}e shows $g_\mathrm{0}$, calculated as in \cite{Schliesser2010B}, assuming the optical mode is localized at the edge of the toroid with negligible transverse size. At the anticrossings, $g_\mathrm{0}$ is reduced (i.e., the effective mass is increased), as a significant part of the mode's energy is in that case associated with displacements that do not modulate the cavity length. Away from the anticrossings, however, the RBM mode is localized exclusively in the toroid and outermost part of the membrane, well isolated from the inner disk and pillar support. As a result, $m_\mathrm{eff}$ is nearly identical to the physical mass of this volume. It is therefore important to minimize the volume of the outermost membrane, i.e., the distance between the spokes and the toroid. As can be seen in Fig. \ref{fig:SampleDesign}f, this simultaneously allows to reach the smallest possible resonance frequency. In practice, the laser reflow process used to form the toroid poses a lower limit on the remaining distance between spokes and toroid.

\subsection{Sample fabrication}

\begin{figure}[]
	\centering
		\includegraphics[width=.5\linewidth]{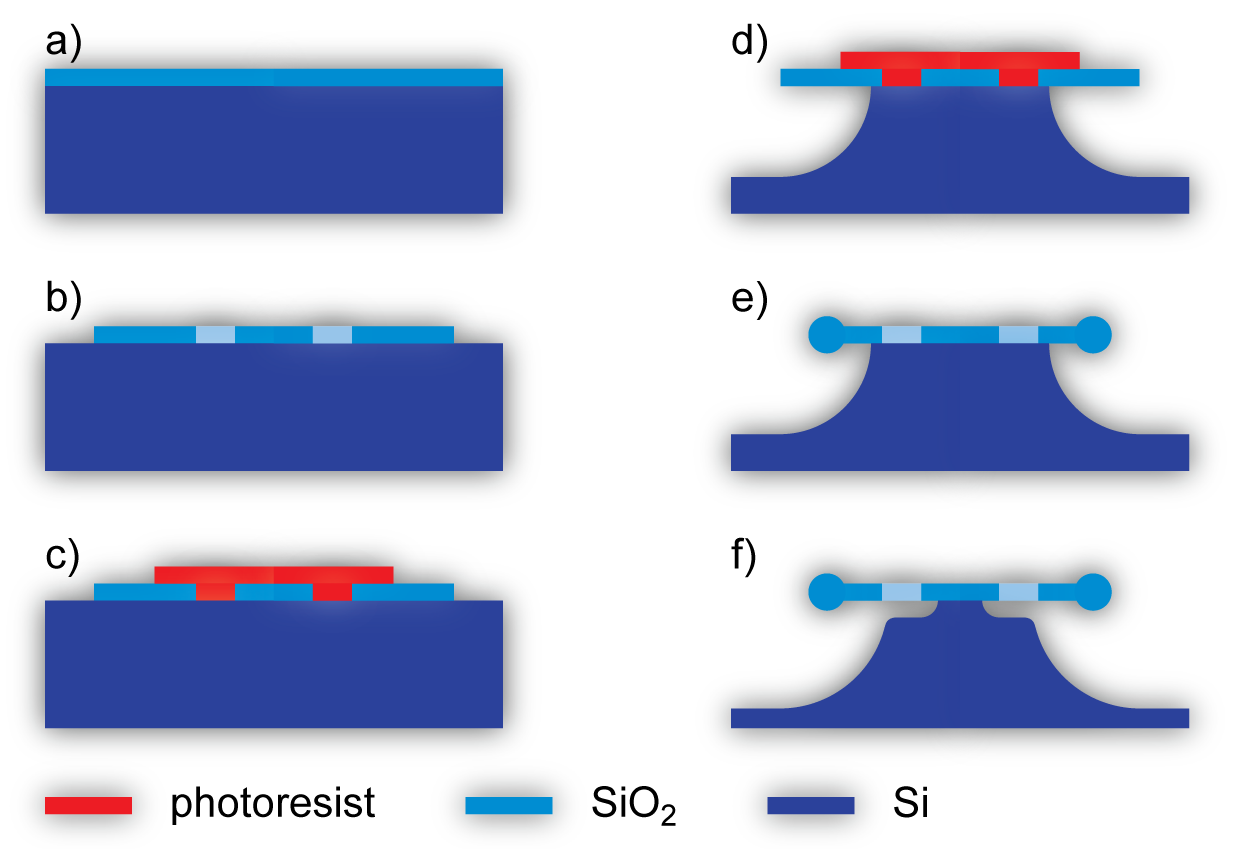}
	\caption{\textbf{Sample fabrication.} See text for details.}
	\label{fig:SampleFabrication}
\end{figure}

To fabricate the spoke-anchored microresonators, we use a combination of optical lithography and dry etching techniques outlined in Fig. \ref{fig:SampleFabrication}. In a first step (b), a disk including the spokes is transferred in a 1~$\mu$m thick film of thermal oxide on a Si wafer (a), through optical lithography followed by reactive ion etching of the SiO$_2$. In a second photolithography step (c), smaller disks of photoresist are defined that cover the center of the SiO$_2$ disks, including the spokes. These serve to protect the exposed Si surface between the spokes during the subsequent isotropic XeF$_2$ etch (d) of the Si substrate. Care is taken to stop the etch shortly before it reaches the apertures in the SiO$_2$ disk. After removing the protective photoresist disks, a laser reflow of the underetched disk is performed (e), forming the silica toroid. Finally (f), a second XeF$_2$ etch releases the toroid and reduces the pillar diameter, typically to a value smaller than 1 $\mu$m.

\subsection{Sample characterization}
The vacuum optomechanical coupling rate $\vcr$ is measured at room temperature in a vacuum chamber. Therefore, the mechanical motion is read out using an external cavity tunable diode laser at 1550 nm, that is locked to a cavity resonance. In order to avoid any radiation pressure effects we perform these measurements at very low laser power (typically around 100 nW). The transmitted light is amplified by a low noise erbium-doped fiber amplifier and sent onto a photodector. For absolute calibration of the mechanical spectrum registered by an electronic spectrum analyzer, we use a phase-modulation technique \cite{Gorodetsky2010B}.
We extract a vacuum optomechanical coupling rate of $\vcr=1700\,\mathrm{Hz}$ for a wavelength of 1550 nm (i.e. $\vcr=3400\,\mathrm{Hz}$ at 780 nm).

\begin{figure}[!b]
	\centering
		\includegraphics[width=0.7\textwidth]{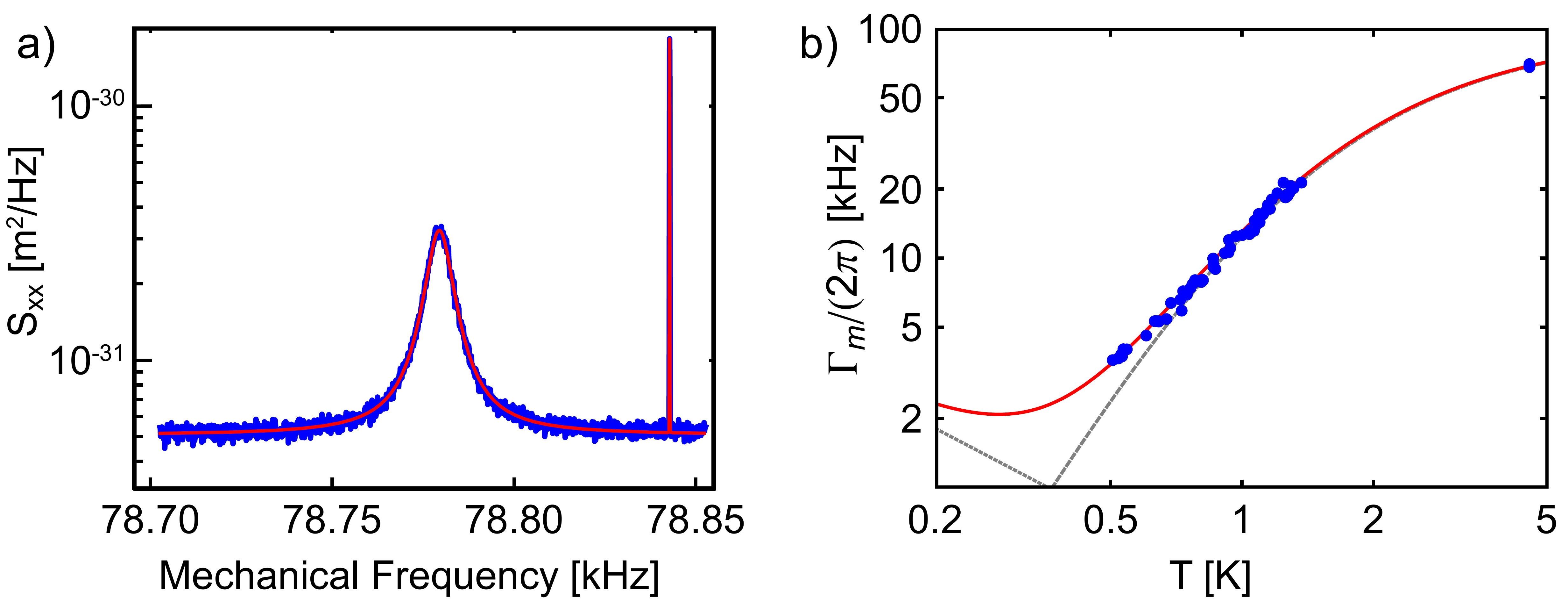}
	\caption{\textbf{Sample characterization.} a) shows a calibrated mechanical noise spectrum for the resonator used throughout the manuscript (except the last panel of Fig. 4 of the main manuscript), which was measured at room temperature in vacuum. The fit (red line) was used to extract a vacuum optomechanical coupling rate of 3400 Hz at 780 nm. b) mechanical damping of this toroid vs. cryostat temperature. The red line is a fit according to the TLS model presented in \cite{Enss2005B,Riviere2011B}, the grey lines represent the contributions from resonant (dotted) and relaxational (dashed) processes. The fit yields a negligible contribution of clamping losses.}
	\label{fig:SampleCharacterization}
\end{figure}

The mechanical linewidth measured at room temperature (8.1 kHz) was found to be higher than expected from the calculated F-parameter (cf. section \ref{SampleDesign}). However, performing the same measurements in the cryostat (cf. Fig. \ref{fig:SampleCharacterization}b), a linewidth as low as 3.6 kHz was found on the same microresonator, indicating that a loss mechanism other than clamping losses must dominate at room temperature. Since there losses due to two level fluctuators (TLS, \cite{Enss2005B,Arcizet2009aB}) have been found to be significantly lower (linewidths below 4 kHz have been measured at room temperature for conventional toroids of similar frequency), we believe that the dominating loss mechanism is thermo-elastic damping (TED) \cite{Nowacki1975B}.
At low temperatures, where TED is strongly reduced, the main loss mechanism is coupling to TLS. Figure \ref{fig:SampleCharacterization}b shows the measured temperature dependence of the mechanical linewidth at low temperature, obtained with the laser (with 100 nW power) resonant with the, in this case, strongly overcoupled optical resonance to avoid dynamical backaction. The variation of $\Gamma_\mathrm{m}$ with temperature can be fitted using a model for the TLS losses \cite{Enss2005B,Riviere2011B}. It is found that this mechanism dominates the total losses for all reachable cryogenic temperatures. This means that it is not possible to retrieve an accurate estimation of the temperature-independent contribution $\Gamma_\mathrm{clamp}$. We can however conclude that it must be at least smaller than 2 kHz for this sample. This shows that in our optimized spoke-supported design, we have successfully mitigated the clamping losses to the level where they are insignificant compared to intrinsic dissipation.

\section{Modeling of optomechanical interaction}

This section summarizes the theoretical model which was used to extract all entities of interest from our data.
Figure \ref{f:model} shows the parameters and variables of the model, and their mutual connections.

\subsection{Conservative dynamics}
The conservative dynamics of an optomechanical system are described by the Hamiltonian \cite{Law1995B}

\begin{align}
   \label{e:cons}
   H&=\frac{1}{4}\hbar \Omega_\mathrm{m}\left(\hq^2+\hmom^2\right)+\hbar \oC \left(\had \ha+\frac{1}{2}\right)+\hbar \vcr \hq \,\had \ha,
\end{align}
where mechanical quadrature operators $q$ and $p$ are related to the corresponding mechanical ladder operators $b$ and $b^\dagger$ via
\begin{align}
  \hq&=b+b^\dagger\\
  \hp	&=(b-b^\dagger)/i.
\end{align}
With these definitions $[\hq,\hp]=2i$, and the actual mechanical displacement and momentum are given by $x'=\xzpf q$ and $p'=\hbar p/2\xzpf$ with the amplitude of the zero-point motion
\begin{align}
  \xzpf&=\sqrt{\frac{\hbar}{2 \meff \Om}}.
\end{align}
The vacuum optomechanical coupling rate $\vcr$ quantifies the strength of the optomechanical interaction and is given by $\vcr=\dwdx \xzpf$ with $\dwdx=\partial \omega_\mathrm{c}/\partial x$.

{\small \begin{figure}[]
\begin{center}
{\small \includegraphics[width=0.85\linewidth]{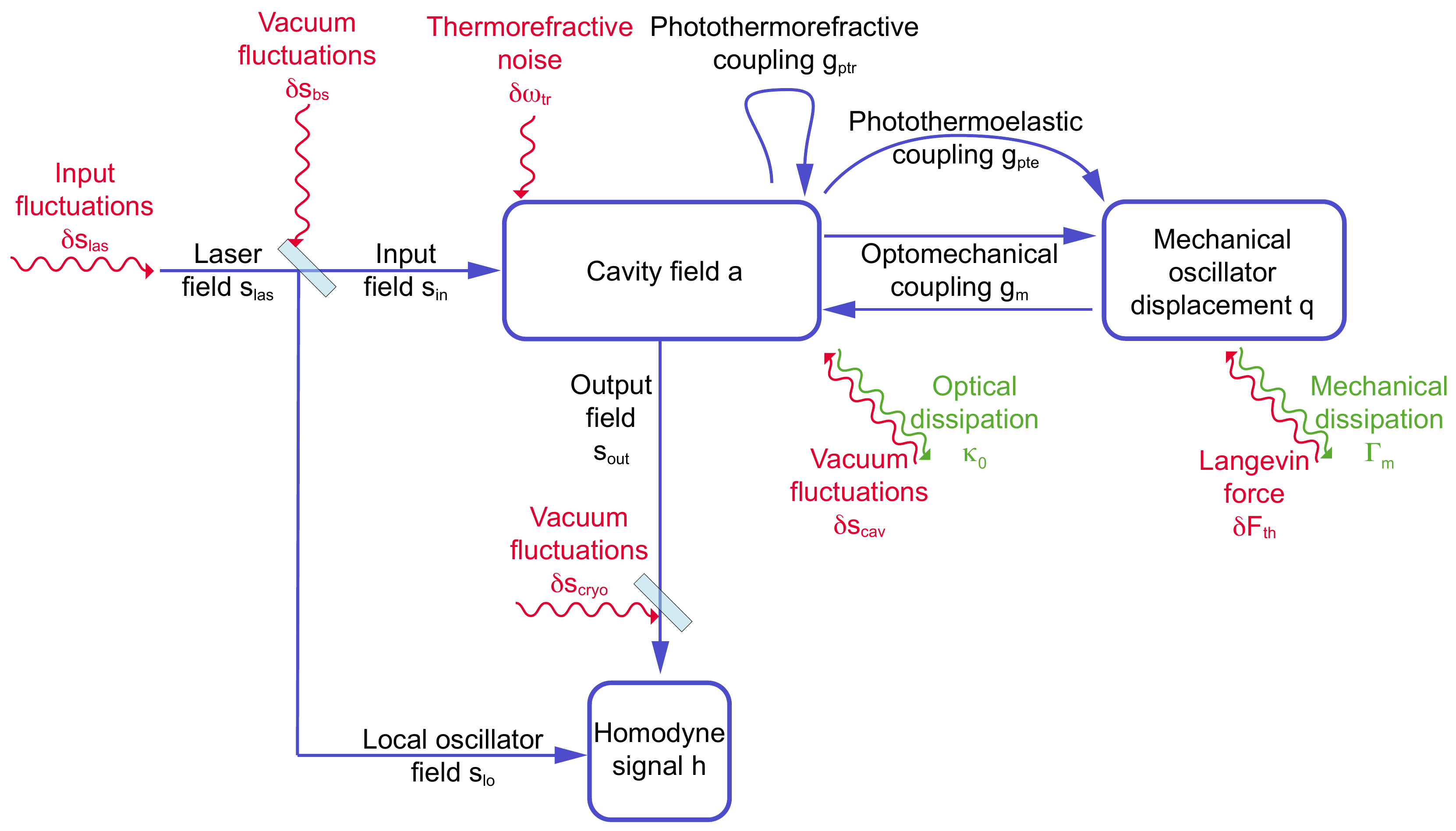}  }
\end{center}
\caption{\textbf{Theoretical model used.} See text for details.}
\label{f:model}
\end{figure}}

\subsection{Quantum Langevin equations}

%\subsubsection{Thermal Langevin force and mechanical dissipation}

The Hamiltonian (\ref{e:cons}) determines the conservative evolution of the optomechanical degrees of freedom.
Optical and mechanical dissipation, and the corresponding fluctuations, can be taken into account by introducing the mechanical dissipation rate $\Gm$ and the optical dissipation rate $\kappa=\kin+\kex$ (where $\kex$ represents losses to the coupling waveguide and $\kin$ all other optical losses) as well as the optical noise terms $\dhsin$, $\dhsvac$ and the thermal Langevin force, which we express as  a rate $\dhFth$  by writing the physical force in momentum units of $\hbar/2\xzpf$.
This leads to the well-known Langevin equations of cavity optomechanics \cite{Fabre1994B, Mancini1994B, Gardiner2004B}
\begin{align}
  \dot\ha(t)&=\left(i \Delta-\frac{\kappa}{2}\right) \ha(t)-  i \vcr \hq(t) \ha(t) +
%  \nonumber\\ \qquad  &+
  \sqrt{\kex} (\bsin+\dhsin(t))+{\sqrt{\kin}}{\dhsvac(t)}\\
  \dot  \hq(t)&=\Om \hp(t)\label{e:pQLE}
%\\
%  \frac{d}{dt}\hp(t)&=-\meff \Om^2 \hx(t)+ \hbar |\dwdx| \had(t) \ha(t)-{\Gm} \hp(t)
 %   + \dhFth(t)
\\
  \dot\hp(t)&=- \Om \hq(t)- 2\vcr  \had(t) \ha(t)-{\Gm} \hp(t)
    + \dhFth(t)
  \label{e:xQLE},
\end{align}
where the convention  $\Delta=\oL-\oC$ was used to denote the detuning of the laser (angular) frequency   $\oL$ from the bare cavity resonance frequency $\oC$, and
$\ha$ is expressed in a frame rotating at $\oL$.

In order to accurately model the response of the optomechanical system over a wide range of parameters (detuning, Fourier frequency, optical and mechanical excitation) for a single set of parameters, we have refined this generic model by including other effects which are known to be inherent to most optical microcavities, and are discussed in the following.

\emph{Photothermoelastic backaction.}
Thermoelastic forces driven by temperature gradients induced by light absorption can induce mechanical displacements.
%
%thermoelastic equations
%\begin{align}
% \mu\, {\mvec \nabla}^2 \mvec u +  (\lambda+\mu) \mvec \nabla ( \mvec \nabla \cdot {\mvec u})+\mvec f &= (3\lambda+2\mu)\alpha \mvec \nabla \theta +\rho {\ddot {\mvec u}}\\
% \label{e:thermaldiffusion}
% k_\mathrm{t} {\mvec \nabla}^2 \theta - c_p \rho \dot \theta&= (3 \lambda+2\mu)\alpha T_0 \mvec \nabla \cdot \dot{ \mvec u}
% 	- \kappa_\mathrm{abs}|a|^2 v^2
%\end{align}
The starting point to model these displacements are the coupled equations of motion
known from the standard theory of thermoelasticity  \cite{Nowacki1975B}
\begin{align}
 \mu\, {\mvec \nabla}^2 \mvec u +  (\lambda+\mu) \mvec \nabla ( \mvec \nabla \cdot {\mvec u})+\mvec f &= (3\lambda+2\mu)\alpha \mvec \nabla \theta +\rho {\ddot {\mvec u}}\\
 \label{e:thermaldiffusion}
 k_\mathrm{t} {\mvec \nabla}^2 \theta - c_\mathrm{t} \rho \dot \theta&= (3 \lambda+2\mu)\alpha T_0 (\mvec \nabla \cdot \dot{ \mvec u}) - v\, \kappa_\mathrm{abs}
 \had \ha.
\end{align}
These equations connect the displacement field $\mvec u (\mvec r,t)$ and the temperature elevation $\theta (\mvec r,t)$ above the mean temperature $T_0$.
Here, $\lambda$ and $\mu$ are the Lam\'e parameters, $\alpha$ the thermal expansion coefficient, $\rho$ the mass density, $\mvec f$ a body force (e.g.\ due to radiation pressure), $k_\mathrm{t}$ the thermal conductivity, $\kappa_\mathrm{abs}$ the photon absorption rate, $c_\mathrm{t}$ the heat capacity, and the function $v(\mvec r)$ describes the spatial distribution of the light absorption.
Evidently, a thermoelastic body force
\begin{align}
 \mvec f_\mathrm{te}(\mvec r,t)=-(3\lambda+2\mu)\alpha \mvec \nabla \theta (\mvec r,t)
\end{align}
acts on the mechanical modes of the structure when a temperature gradient $\mvec\nabla \theta (\mvec r,t)$ is present.
In the scalar representation of the mechanical dynamics, we therefore have to add a thermoelastic force $f_\mathrm{te}(t)$ proportional to the material parameters $\lambda$, $\mu$ and $\alpha$, as well as an overlap integral of the mechanical mode's displacement pattern and the temperature gradient $\mvec \nabla \theta (\mvec r,t)$.
Assuming that the temperature gradients are predominantly driven by the absorption of laser light in the resonator%
%(second term in RHS of eq.\ \ref{
, one can express the scalar photothermoelastic force as
\begin{align}
  f_\mathrm{pte}(t)= \chi_\mathrm{pte}(t)\ast�\kappa_\mathrm{abs} a^\dagger(t) a(t),
\end{align}
where we have absorbed the spatial overlap integrals between the mechanical and (the gradient of) the thermal modes, as well as the thermal modes and the spatial pattern of light absorption into the magnitude of the function $\chi_\mathrm{pte}(t)$.
The temporal dynamics of the adjustment of the relevant temperature gradients to a changing amount of light absorption is represented by the time-dependence of $\chi_\mathrm{pte}(t)$ (``$\ast$'' denotes a convolution).
Note that while this formulation accounts for the quantum fluctuations of the intracavity field $a(t)$, the statistical nature of photon absorption events is neglected.
This is justified considering that the quantum fluctuations of optical heat deposition (``photothermoelastic shot noise'') have a much smaller effect on the mechanical mode than the direct fluctuations of the radiation pressure term $2\vcr\, a^\dagger a$.

\emph{Dynamic photothermorefractive frequency shift.}
A temperature elevation $\theta(\mvec r,t)$ within the optical mode volume furthermore changes the refractive index, and therefore the optical resonance frequency.
In analogy to the description in the previous section, we are using a simple scalar description of the form
\begin{align}
  \Delta \omega_\mathrm{ptr}(t)= \chi_\mathrm{ptr}(t)\ast\kappa_\mathrm{abs} a^\dagger(t) a(t)
\end{align}
for this frequency shift, where the response function $\chi_\mathrm{ptr}(t)$ accommodates spatial overlap integrals of the light absorption pattern $v(\mvec r)$ and the thermal modes as well as the temporal dynamics of the latter, and, in addition, the spatial sampling of the induced refractive index changes
\begin{align}
  \Delta n(\mvec r,t)=\frac{\mathrm{d}n}{\mathrm{d}T} \theta(\mvec r,t)
\end{align}
by the optical mode.

\emph{Thermorefractive noise.} The local temperature elevation $\theta(\mvec r, t)$ also undergoes {thermal} fluctuations\textemdash independent of the presence of light.
Within a volume $V$, they amount to squared fluctuations of \cite{Landau1980B}
\begin{align}
 \left\langle   \theta(\mvec r, t)^2 \right\rangle_{V}=\frac{k_\mathrm{B} T_0^2}{c_p \rho V}.
\end{align}
The spatial distribution of these fluctuations can be calculated using a Langevin ansatz \cite{Braginsky1999B}, by adding a fluctuational source term to the heat diffusion equation (\ref{e:thermaldiffusion}). Predominantly via the thermorefractive effect ($\mathrm{d}n/\mathrm{d}T\neq 0$), the resulting temperature fluctuations again induce resonance frequency fluctuations $\delta \omega_\mathrm{tr}(t)$.
Its temporal correlation function (or equivalently, power spectral density) have been estimated for simple whispering-gallery mode resonator geometries \cite{ Gorodetsky2004B, Schliesser2008bB, Anetsberger2010B}.
%
%
%\subsubsection{Full model equations}

Taking these additional three effects into account, the equations of motion can be written as
\begin{align}
  \dot\ha(t)&=\left(i (\Delta-\Delta\omega_\mathrm{ptr}(t)-\delta\omega_\mathrm{tr}(t))-\frac{\kappa}{2}\right) \ha(t)- i \vcr \hq(t) \ha(t) +
  \nonumber\\ \qquad  &+
  \sqrt{\kex} (\bsin+\dhsin(t))+{\sqrt{\kin}}{\dhsvac(t)}\\
  \dot\hq(t)&=\Om \hp(t)
%\\
%  \frac{d}{dt}\hp(t)&=-\meff \Om^2 \hx(t)+ \hbar |\dwdx| \had(t) \ha(t)-{\Gm} \hp(t)
 %   + \dhFth(t)
\\
  \dot\hp(t)&=- \Om \hq(t)- 2 \vcr  \had(t) \ha(t)-{\Gm} \hp(t)
    +\dhFth(t)+ f_\mathrm{pte}(t)
    .
\end{align}

\subsection{Linearized model}
A large coherent field sent to the optomechanical system induces a relatively large classical intracavity field $\ba$, and induces a displacement of the mechanical mode by $\bar q$.
If the system is stable around this steady-state, the dynamics of the small fluctuations around this equilibrium are described by a set of equations obtained via the substitution $\ha(t)=\ba+\dha(t)$ and $\mhat q(t)=\bar q+\dhq(t)$, and retaining only first-order terms in the fluctuations.
This yields
\begin{align}
   \dot\dha(t)&=\left(+i \bar\Delta-\frac{\kappa}{2}\right) \dha(t)
   -i  \kappa_\mathrm{abs} \ba \chi_\mathrm{ptr}(t) \ast  (\ba^*  \dha(t)+ \ba \dhad(t))
   - i \vcr \ba \dhq(t)  -
   \nonumber\\  &\qquad -
  i \ba \delta \omega_\mathrm{tr}(t) +\sqrt{\kex} \dhsin(t)+{\sqrt{\kin}}{\dhsvac(t)}
  \end{align}
\begin{align}
 \Om^{-1}\left[ \delta \ddot{ \hq}(t)+{\Gm} \, \delta \dot{\hq}(t)+ \Om^2\, \dhq(t)\right]&=- 2 \vcr (\ba\dhad(t)+\ba^* \dha(t))
    + \nonumber\\ &\qquad
  +\dhFth(t)+\chi_\mathrm{pte}(t)\ast(\ba\dhad(t)+\ba^* \dha(t))
 \end{align}
with $\bar \Delta=\oL-\left(\oC+\vcr\bar q+	\kappa_\mathrm{abs} |\ba|^2 (\chi_\mathrm{ptr}(t)\ast 1)\right)$. This set of equations is best solved in the Fourier domain, yielding
\begin{align}
   \left(-i(\bar\Delta+\Omega)+\kappa/2\right)\dha(\Og)&=
   -i\kappa_\mathrm{abs} \ba \chi_\mathrm{ptr}(\Og)   (\ba^*  \dha(\Og)+ \ba \dhad(\Og))
   -  i \vcr \ba \dhq(\Og)  +
   \nonumber\\  &\qquad
   -i \ba \delta \omega_\mathrm{tr}(\Og) + \sqrt{\kex} \dhsin(\Og)+{\sqrt{\kin}}{\dhsvac(\Og)}
%  \end{align}
%\begin{align}
    %
    \\
  \frac{- \Og^2- i \Og \Gm + \Om^2}{\Om} \dhq(\Og)&=\left(-2 \vcr  +\chi_\mathrm{pte}(\Og) \right)(\ba\dhad(\Og)+\ba^* \dha(\Og))
  %+ \nonumber\\ &\qquad
    + \dhFth(\Og).
 \end{align}
For simplicity, we refer to the Fourier transform of the respective functions by simply writing them with a frequency ($\Omega$) argument. Note that $\dhad(\Og)$ denotes the Fourier transform of $\dhad(t)$, equal to $[\dha(-\Og)]^\dagger$; and that $[\dhq(-\Og)]^\dagger=\dhq(\Og)$ for the Hermitian operator $\dhq(t)$.

To further simplify the problem, we approximate the response functions of the photothermal effects by a single-pole, low-pass response, assuming implicitly that the relevant temperature (gradient) distributions adjust themselves only with a certain delay to a change in the absorbed optical power.
Assuming that this delay is larger than the relevant oscillation periods considered here, one can approximate
\begin{align}
	\chi_\mathrm{ptr}(\Omega)&\approx  \frac{g_\mathrm{ptr}}{\kappa_\mathrm{abs}} \frac{\Om}{-i\Og}�\\
	\chi_\mathrm{pte}(\Omega)&\approx{2 g_\mathrm{pte}}{} \frac{\Om}{-i\Og}
\end{align}
and finally obtains
\begin{align}
   \left(-i(\bar\Delta+\Omega)+\kappa/2\right)\dha(\Og)&=
   \ba {g_\mathrm{ptr}} \frac{\Om}{\Og} (\ba^*  \dha(\Og)+ \ba \dhad(\Og))
   - i \vcr \ba \dhq(\Og)  +
   \nonumber\\  &\qquad
  -i \ba \delta \omega_\mathrm{tr}(\Og) + \sqrt{\kex} \dhsin(\Og)+{\sqrt{\kin}}{\dhsvac(\Og)}
  		\label{e:omsimple}
%  \end{align}
%\begin{align}
    %
    \\
 \frac{- \Og^2- i \Og \Gm + \Om^2}{\Om}  \dhq(\Og)&=-2\left(\vcr  + i g_\mathrm{pte}\frac{\Om}{\Og}\right)(\ba\dhad(\Og)+\ba^* \dha(\Og))
%  + \nonumber\\ &\qquad
    + \dhFth(\Og).
 \end{align}
 These equations are used to calculate the coherent response and fluctuation spectra (cf. sections \ref{NoiseCovariances}, \ref{CoherentDynamics}).

\subsection{Homodyne detection \label{HomodyneDetection}}

The optomechanical experiment is embedded into one arm of a balanced homodyne interferometer. At the initial beamsplitter, the laser field (and fluctuations in the  fiber mode) are split up into a `local oscillator' arm, and the arm that serves as input to the cavity:
\begin{align}
   s_\mathrm{in}&=\sqrt{1-r}s_\mathrm{las}-\sqrt{r}s_\mathrm{bs}\\
   s_\mathrm{lo}&=\sqrt{r}s_\mathrm{las}+\sqrt{1-r}s_\mathrm{bs},
\end{align}
evidently valid both in time and frequency domain. Here, we also take into account the vacuum fluctuations $\delta s_\mathrm{bs}$ entering the beamsplitter at the unoccupied port,
\begin{align}
   s_\mathrm{las}&= \bar s_\mathrm{las}+\delta s_\mathrm{las}\\
   s_\mathrm{bs}&=\delta s_\mathrm{bs}.
\end{align}
The field $s_\mathrm{in}$ drives both the mean field $\ba$ and the field fluctuations within the cavity, as described in the previous section.
The intracavity field $a$, in turn, couples back into the single-mode fiber taper, and the usual input-output formalism gives the field $s_\mathrm{out}$ at the output of the cavity via the relation
\begin{align}
   s_\mathrm{in}-s_\mathrm{out}=\sqrt{\kex} a
\end{align}
We furthermore take into account that only a fraction $\eta_\mathrm{cryo}$ of the light power at the output of the cavity is measured as `signal' in the homodyne detector due to optical losses, e.g.\ in the cryostat.
For $\eta_\mathrm{cryo}< 1$, we again have to account for quantum vacuum $\delta s_\mathrm{cryo}$ that enters the optical mode,
\begin{align}
   s_\mathrm{sig}&=\sqrt{ \eta_\mathrm{cryo}}s_\mathrm{out}+\sqrt{1- \eta_\mathrm{cryo}}s_\mathrm{cryo}\\
   s_\mathrm{cryo}&=\delta s_\mathrm{cryo}.
\end{align}
Finally, in the homodyne receiver, the differential signal
%\begin{align}
%  \delta h&= \bar s_\mathrm{lo} e^{+i \phi_\mathrm{lo}}
%  			 \left(\sqrt{ \eta_\mathrm{cryo}} \delta s_\mathrm{out}^\dagger+\sqrt{ 1- \eta_\mathrm{cryo}} \delta s_\mathrm{cryo}^\dagger \right)
%			+ \bar s_\mathrm{lo}^* e^{-i \phi_\mathrm{lo}}
%  			 \left(\sqrt{ \eta_\mathrm{cryo}} \delta s_\mathrm{out}+\sqrt{ 1- \eta_\mathrm{cryo}} \delta s_\mathrm{cryo} \right)
%			 \nonumber\\ &\qquad
%			+\sqrt{ \eta_\mathrm{cryo}}\bar s_\mathrm{out} e^{-i \phi_\mathrm{lo}} \delta s_\mathrm{lo}^\dagger+\sqrt{ \eta_\mathrm{cryo}}\bar s_\mathrm{out}^* e^{+i \phi_\mathrm{lo}} \delta s_\mathrm{lo}
%		\label{e:homo}
%\end{align}
\begin{align}
  \delta h&=    \bar s_\mathrm{lo} e^{+i \phi_\mathrm{lo}} \delta s_\mathrm{sig}^\dagger
			+ \bar s_\mathrm{lo}^* e^{-i \phi_\mathrm{lo}} \delta s_\mathrm{sig}
%			 \nonumber\\ &\qquad
			+\bar s_\mathrm{sig} e^{-i \phi_\mathrm{lo}} \delta s_\mathrm{lo}^\dagger+\bar s_\mathrm{sig}^* e^{+i \phi_\mathrm{lo}} \delta s_\mathrm{lo}
			\label{e:homo}
\end{align}
is measured.
The fluctuational terms $\delta h$ and $\dhq$ of interest can then be expressed as a linear function of
the fluctuations driving the system,
\begin{align}
  \begin{pmatrix} \delta h \\  \dhq  \end{pmatrix}
  &=M\cdot
    \begin{pmatrix}
  \delta s_\mathrm{las}&
  \delta s_\mathrm{las}^\dagger&
  \delta s_\mathrm{bs}&
  \delta s_\mathrm{bs}^\dagger&
  \dhsvac&
  \dhsvacd&
  \delta s_\mathrm{cryo}&
  \delta s_\mathrm{cryo}^\dagger&
  \delta \omega_\mathrm{tr}&
  \delta\! f_\mathrm{th}
  \end{pmatrix}^T.
  \label{e:aux1}
\end{align}
Here, the coefficients of the matrix $M$ follow directly from the relations (\ref{e:omsimple})-(\ref{e:homo}).

\subsection{Calculation of noise covariances \label{NoiseCovariances}}
\label{ss:covariances}

We assume that all input noise terms of eq.\ (\ref{e:aux1}) can be described by zero-mean Gaussian noise operators whose variances are known.
Representing the covariances between two noise operators $x$ and $y$ as a symmetrized spectrum $\bar S_{xy}(\Omega)$ defined according to
%\begin{align}
%  \frac{1}{2} \left \langle \left \{ \dhsin(\Og), \dhsind (\Og') \right\} \right\rangle=2\pi \bar S_{\dhsin,\dhsind}(\Og)\,\delta(\Og+\Og')
%\end{align}
\begin{align}
  \frac{1}{2} \left \langle \left \{ x(\Og),y(\Og') \right\} \right\rangle=2\pi \bar S_{x y}(\Og)\,\delta(\Og+\Og'),
\end{align}
%\begin{align}
%	\bar S_{\dhsind \dhsin}(\Og)&=\frac{1}{2}\\
%	\bar S_{\dhsvacd \dhsvac}(\Og)&=\frac{1}{2}\\
%	\bar S_{ \delta s_\mathrm{bs}^\dagger  \delta s_\mathrm{bs}}(\Og)&=\frac{1}{2}\\
%	\bar S_{ \delta s_\mathrm{cryo}^\dagger \delta s_\mathrm{cryo}}(\Og)&=\frac{1}{2}\\
%	\bar S_{ \delta F_{th} \delta F_\mathrm{th}}(\Og)&=\hbar \meff \Gm \Og \coth \left(\frac{\hbar \Omega}{2 k_\mathrm{B} T} \right)
%				\approx \frac{\hbar^2}{\xzpf^2}\Gm n_\mathrm{i}\\
%	\bar S_{ \delta \omega_{tr} \delta \omega_\mathrm{tr}}(\Og)&=\frac{1}{}
%\end{align}
the only non-zero covariances are characterized by the spectra
\begin{align}
	\bar S_{ \delta s_\mathrm{las}^\dagger  \delta s_\mathrm{las}}(\Og)&=
	\bar S_{\dhsvacd \dhsvac}(\Og)=
	\bar S_{ \delta s_\mathrm{bs}^\dagger  \delta s_\mathrm{bs}}(\Og)=
	\bar S_{ \delta s_\mathrm{cryo}^\dagger \delta s_\mathrm{cryo}}(\Og)=\frac{1}{2}
	\label{e:qn}
\end{align}
	for the optical quantum noise entering the system,
\begin{align}
	\bar S_{ \dhFth \dhFth}(\Og)&\approx 4  \bar n_\mathrm{m} \Gm
	\label{e:tn}
\end{align}
	{for the thermal Langevin force, where we have assumed $ \bar n_\mathrm{m}\approx k_\mathrm{B} T/\hbar \Om\gg 1$, and}
\begin{align}
	\bar S_{ \delta \omega_ \mathrm{tr} \delta \omega_\mathrm{tr}}(\Og)&=\bar S_\mathrm{trn}(\Og),
	\label{e:trn}
\end{align}
for the thermorefractive noise \cite{Gorodetsky2004B}, whose contribution we found to be negligible in the data presented in this manuscript.
By the linearity of equation (\ref{e:aux1}), it follows that the covariance matrix $N_\mathrm{out}$ of the output noise operators is then related to the input covariance matrix $N_\mathrm{in}$ by the simple expression
\begin{align}
	N_\mathrm{out}&=M(+\Og)\cdot N_\mathrm{in} \cdot M(-\Og)^T.
	\label{e:noise}
\end{align}

\subsection{Coherent dynamics of the system \label{CoherentDynamics}}

In order to calculate the coherent response of the system to the probing by a phase-modulated input, eq.\ (\ref{e:aux1}) can be used. By assuming a sufficiently narrow detection bandwidth and/or sufficiently large phase modulation of depth $\delta \varphi$, one can set
\begin{align}	
%	0& \approx \dhsvacd(\Og)\approx\dhsvac(\Og)\approx \delta s_\mathrm{bs}^\dagger(\Og) \approx \delta s_\mathrm{bs}(\Og)\approx\nonumber\\
%	&\approx \delta s_\mathrm{cryo}^\dagger(\Og) \approx\delta s_\mathrm{cryo}(\Og)\approx \dhFth(\Og)\approx \delta \omega_{tr}(\Og)\\
	\dhsvacd&\approx\dhsvac\approx \delta s_\mathrm{bs}^\dagger \approx \delta s_\mathrm{bs}\approx\delta s_\mathrm{cryo}^\dagger
 	\approx\delta s_\mathrm{cryo}\approx \dhFth\approx \delta \omega_{tr}\approx 0
	\label{e:coh1}\\
	\delta s_\mathrm{las}&= i \bar s_\mathrm{las} \delta \varphi,
	\label{e:coh2}
\end{align}
and calculate the frequency-dependent transfer function  from a phase modulation $\delta \varphi$ to the homodyne signal $\delta h$.
%\begin{align}	
%	\bar S_{\dhsvacd \dhsvac}(\Og)&=
%	\bar S_{ \delta s_\mathrm{bs}^\dagger  \delta s_\mathrm{bs}}(\Og)=
%	\bar S_{ \delta s_\mathrm{cryo}^\dagger \delta s_\mathrm{cryo}}(\Og)=
%	\bar S_{ \delta F_{th} \delta F_\mathrm{th}}(\Og)=
%	\bar S_{ \delta \omega_{tr} \delta \omega_\mathrm{tr}}(\Og)=0
%\end{align}

This coherent response is obviously directly measured in the sideband sweeps that we routinely perform (cf.\ section \ref{ss:setup}).
Moreover, by multiplication of the (complex) spectrum of the excitation pulse with this transfer function, the response of the homodyne signal in the time domain can be numerically determined via the inverse Fourier transform.

\subsection{Analysis of the coherent response}
The coherent response spectra are important to accurately extract the different parameters of the optomechanical interaction as well as to calibrate the mechanical noise spectra.
A typical coherent response is shown in Fig. \ref{f:CoherentResponse}a. The Lorentzian peak centered around $\Omega_\mathrm{mod}=140$~MHz results from the absorption of the upper modulation sideband by the cavity and reflects the optical response of the system. The maximum of the homodyne signal is obtained when the modulation sideband is resonant with the cavity. Hence, the center frequency and width of this peak correspond to the detuning $|\Delta|$ and the linewidth $\kappa$ of the cavity, respectively.
The sharp feature at $\Omega_\mathrm{mod} = \Omega_\mathrm{m}$ is the manifestation of Optomechanically Induced Transparency \cite{Weis2010B}; an interference effect due to the resonant excitation of the mechanical mode.
For weak coupling power and/or large detuning, the dynamics of the mechanical mode is hardly affected by the optomechanical interaction and the width of the dispersive feature is given by the mechanical linewidth $\Gm$.

For larger laser power, the width of the OMIT window increases, reflecting the width of the damped mode  $\Gamma_\mathrm{m} + \Omega_c^2 \kappa / \left( \kappa^2 + 4 (\Delta + \Omega_\mathrm{m})^2 \right)$. Hence, the fit of the coherent response allows to extract the coupling rate $\Omega_c$ and the corresponding intracavity field $\bar a$. We introduce $\ba_0\equiv\ba/\frac{\kappa/2}{-i \bD+\kappa/2}$ to obtain a parameter independent of detuning.
The model of eq.\ (\ref{e:aux1}), assuming pure radiation pressure backaction, fits the measurements well (cf.\ Fig.~\ref{f:CoherentResponse}a).
However, as can be seen in Fig.~\ref{f:CoherentResponse}b, a small systematic deviation appears for high coupling power. This systematic effect is very well reproduced by the model including the photothermoelastic effect.

Finally, Figure \ref{f:CoherentResponse}c shows a series of coherent response spectra taken for decreasing laser detunings, and  a laser power of 0.6 mW.
The observed increase of the amplitudes for small detuning can be fitted accurately by introducing the photothermorefractive effect in the model (red lines). The parameter $g_\mathrm{ptr}$ introduced here is dependent on detuning since the thermorefractive coefficient $\frac{\mathrm{d}n}{\mathrm{d}T}$ depends on temperature.

For example, we have extracted from the fits to the full detuning series of Fig.\ 2 of the main manuscript
$\Om/2\pi=78.2\unit{MHz}$,
$\kappa/2\pi=6.0\unit{MHz}$,
$\ba_0=14.2\cdot10^3$ (with $\vcr/2\pi=3.4\unit{kHz}$),
$g_\mathrm{pte}/2\pi=-122\unit{Hz}$
and $g_\mathrm{ptr}/2\pi=0.32\unit{Hz}$ (at the lower mechanical sideband).

{\small \begin{figure}[]
\begin{center}
{\small \includegraphics[width=\linewidth]{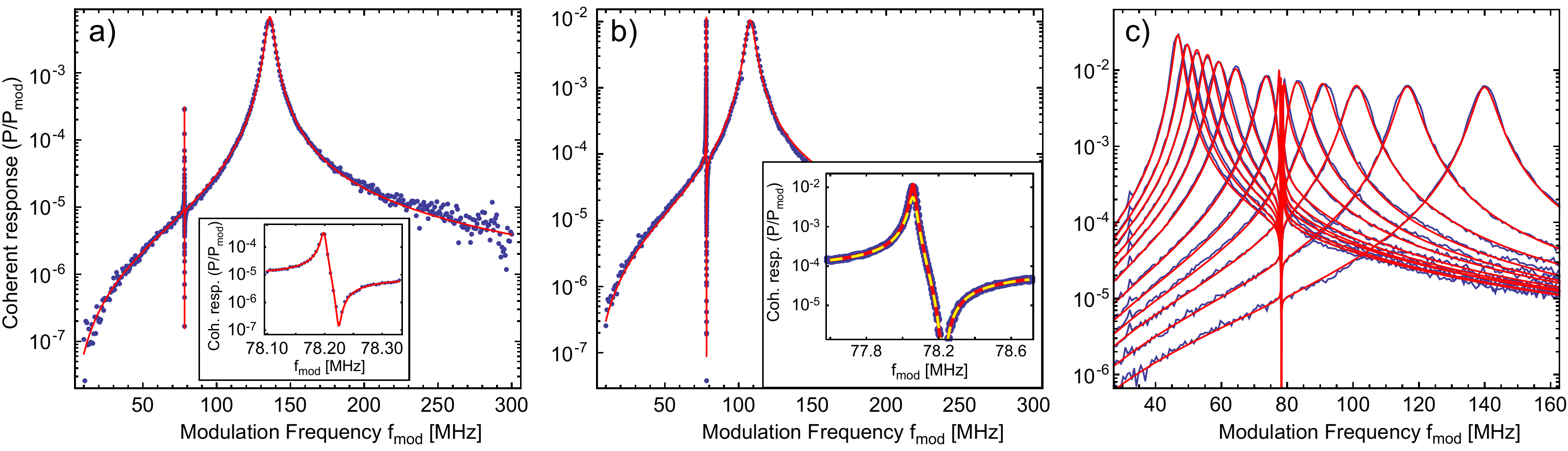}  }
\end{center}
\caption{\textbf{Fitting the model to the coherent response.} (a) A coherent response spectrum taken with a power of 0.56~mW, at T=0.65~K. (b) Spectrum for 1.4~mW, at T=0.8~K with fits including the photothermoelastic effect (red line) and without (yellow dashed).
(c) Spectra for 0.6~mW at T=0.75~K, for various detunings. The photothermorefractive effect is included in the fitted model and accounts for the increased amplitude for small detuning.
}
\label{f:CoherentResponse}
\end{figure}}

\subsection{Extraction of the decoherence rate}

The fits of the coherent response spectra determine all parameters characterizing the optomechanical interaction, and therefore the transduction of mechanical displacement fluctuations to optical fluctuations.
The spectral shape of the noise originating from the Langevin force is thus fixed, so that the amplitude of this contribution can be fitted using the model of
eq.\ (\ref{e:noise}).
As we fit the spectral density of the actually measured \emph{voltage} signal, these  extracted amplitudes depend on the gain of the subsequent detection chain, which is not precisely known.

This ambiguity is removed by a calibration technique \cite{Riviere2011B} based on a reference phase modulation, which allows to relate noise spectra taken under arbitrarily different acquisition conditions.
In this manner, we link the low-temperature noise spectra to a measurement at a higher cryostat temperature (4 K), in which a high helium gas pressure, and low optical power ($\sim100\,\mathrm{nW}$) ensure the thermalization of the sample, so that the Langevin force is known to an estimated accuracy of 3\%.
In this high-temperature measurement,  a known phase modulation is applied, whose amplitude can be compared with the coherent response spectra acquired with every low-temperature measurement.
Assuming that no drift occurs in the phase modulation chain, this method allows to absolutely calibrate the Langevin force\textemdash and therefore the mechanical decoherence rate\textemdash in the low-temperature measurements.
Importantly, this derivation reveals possible changes of the decoherence rate both due to a changed temperature (bath occupation $\bar n_\mathrm{m}$) and mechanical dissipation rate $\Gm$.

\subsection{Error analysis \label{ErrorAnalysis}}
We use the large number of traces acquired during a detuning sweep to estimate an error on each of the four parameters assumed to be independent of the detuning ($\Omega_\mathrm{m}$, $\kappa$, $\bar{a}_\mathrm{0}$, $g_\mathrm{pte}$). This is achieved by successively letting each of these parameters vary with the detuning, while the three others are still fitted globally. The error on each parameter $X$ is obtained by calculating the standard deviation $
\Delta X = \sqrt{\langle(X-X_\mathrm{0})^2\rangle},
$
where $X_\mathrm{0}$ is the value obtained when all four parameters are kept constant over the whole detuning range. Advantageously, this procedure reflects also systematic errors due to drifts of the experimental settings over the detuning series, and physical effects that are not captured by the model.
The following uncertainties were obtained with this method for the run presented in Fig. 2 of the main manuscript:
$ \Omega_\mathrm{m}/2\pi=(78.2260\pm0.0007)\unit{MHz}$, $\kappa/2\pi =(6.04\pm0.08)\unit{MHz}$,
$\ba_{0}  = (14.2\pm0.2)\times 10^3$, $g_\mathrm{pte}/2\pi=(122\pm52)\unit{Hz}$.
These errors, by affecting the shape of the expected noise spectra, also translate in an error on the fitted decoherence rate and occupation. A Monte-Carlo approach is used to assess the final error on $\gamma$ and $\bar{n}$. The fit of the noise spectrum is repeated with a set of  randomly drawn parameters, assuming an independent normal distribution for each of the previous parameters. Importantly, the resulting uncertainty depends on the particular detuning point. On the lower optomechanical sideband, the standard deviation of the results is given by $\left(\Delta \gamma/\gamma\right)_\mathrm{model} = 6 \,\%$ and $\left(\Delta \bar n/\bar n\right)_\mathrm{model} =4\,\%$.
Another source of uncertainty for these two parameters is the independent calibration of the optomechanical transduction that we estimate to be on the order of $\Delta_\mathrm{calib} = 3 \%$ from the scatter between calibration measurements taken at different probing power. Finally, as discussed in section \ref{GAWBS}, an uncertainty $\Delta_\mathrm{GAWBS}$ is quadratically added to account for the possible presence of GAWBS in the optical fibers before the cavity. The total error for this example is given by
\begin{align*}
\frac{\Delta \gamma}{\gamma} = \sqrt{{\left(\frac{\Delta \gamma} {\gamma}\right)^2}_\mathrm{model} + {\Delta_\mathrm{calib}}^2  + \Delta_\mathrm{GAWBS}^2} = 10\,\%\\
\frac{\Delta \bar n}{\bar n} = \sqrt{{\left(\frac{\Delta n} {n}\right)^2}_\mathrm{model} + {\Delta_\mathrm{calib}}^2  + \Delta_\mathrm{GAWBS}^2} = 7\,\%.
\end{align*}

%\bibliographystyle{plain}
%\bibliography{microCavities}
%\bibliographystyle{Nature}
%\bibliography{verhagenReport}

\end{document}